\documentclass{article}

\usepackage{times}

\usepackage{llncsdoc}
\usepackage{amsmath}
\usepackage{amssymb}
\usepackage{latexsym}
\usepackage{algorithm}
\usepackage{algorithmic}
\usepackage[dvips]{graphicx,color}
\usepackage{setspace}

\usepackage{amsfonts}
\usepackage{enumerate}

\newtheorem{Corollary}{Corollary}

\newtheorem{lemma}{Lemma}
\newtheorem{proposition}{Proposition}

\newtheorem{theorem}{Theorem}
\newtheorem{defn}{Definition}

\newcommand{\ti}[1]{\textit{#1}}
\newcommand{\tbf}[1]{\textbf{#1}}
\newcommand{\mc}[1]{\mathcal{#1}}

\newcommand{\Blacksquare}{\rule{.7em}{.7em}}
\newcommand{\epf}{\hfill $\Blacksquare$}

\newcommand{\lt}{{{\L}o{\'s}-Tarski}}
\newcommand{\xset}{\{x_1,\ldots,x_k\}}
\newcommand{\aset}{\{a_1,\ldots,a_k\}}
\newcommand{\M}[2]{\ensuremath{{#1}({#2})}}
\newcommand{\PS}{\ensuremath{\mathbb{PS}}}
\newcommand{\PSC}{\ensuremath{\mathbb{PSC}}}
\newcommand{\PSCf}{\ensuremath{\mathbb{PSC}_f}}

\begin{document}

\title{Preservation under Substructures modulo Bounded Cores}

\author{Abhisekh Sankaran, Bharat Adsul, Vivek Madan,\\ Pritish
  Kamath, Supratik Chakraborty}
\date{\small{Indian Institute of Technology (IIT), Bombay, India}\\
\footnotesize{\texttt{\{abhisekh, adsul, vivekmadan, pritishkamath, supratik\}@cse.iitb.ac.in}}}
\maketitle
\thispagestyle{empty}

\begin{abstract}
\noindent We investigate a model-theoretic property that generalizes
the classical notion of ``preservation under substructures''. We call
this property \emph{preservation under substructures modulo bounded
  cores}, and present a syntactic characterization via $\Sigma_2^0$
sentences for properties of arbitrary structures definable by FO
sentences.  As a sharper characterization, we further show that the
count of existential quantifiers in the $\Sigma_2^0$ sentence equals
the size of the smallest bounded core. We also present our results on
the sharper characterization for special fragments of FO and also over
special classes of structures. We present a (not FO-definable) class
of finite structures for which the sharper characterization fails, but
for which the classical {\lt} preservation theorem holds. As a fallout
of our studies, we obtain combinatorial proofs of the {\lt} theorem
for some of the aforementioned cases.
\end{abstract}


\section{Introduction}
Preservation theorems have traditionally been an important area of
study in model theory. These theorems provide syntactic
characterizations of semantic properties that are preserved under
model-theoretic operations. One of the earliest preservation theorems
is the {\lt} theorem, which states that over arbitrary structures, a
First Order (FO) sentence is preserved under taking substructures iff
it is equivalent to a $\Pi^0_1$ sentence ~\cite{chang-keisler}.
Subsequently many other preservation theorems were studied,
e.g. preservation under unions of chains, homomorphisms, direct
products, etc.  With the advent of finite model theory, the question
of whether these theorems hold over finite structures became
interesting. It turned out that several preservation theorems fail in
the finite ~\cite{gurevich-alechina, gurevich-shelah, rosen}.  This
inspired research on preservation theorems over special classes of
finite structures, e.g. those with bounded degree, bounded treewidth
etc.  These efforts eventually led to some preservation theorems being
``recovered'' ~\cite{dawar-pres-under-ext, dawar-hom}.  Among the
theorems whose status over the class of all finite structures was open
for long was the homomorphism preservation theorem. This was recently
resolved in ~\cite{rossman-hom}, which showed that the theorem
survives in the finite.

In this paper, we look at a generalization of the `preservation under
substructures' property that we call \ti{preservation under
  substructures modulo bounded cores}. In Section
\ref{section:pres-under-subst-mod-cores}, we show that for FO
sentences, this property has a syntactic characterization in terms of
$\Sigma^0_2$ sentences over arbitrary structures. As a sharper
characterization, we state our result (but provide the proof later in
Section \ref{section:proof-of-conjecture}) that for core sizes bounded
by a number $B$, there is a syntactic characterization in terms of
$\Sigma^0_2$ sentences that use atmost $B$ existential quantifiers.
In Section \ref{section:relativization}, we discuss how the notion of
\emph{relativization} can be used to prove the sharper
characterization in special cases and also discuss its limitations.
We present our studies for special classes of FO and over special
classes of structures in Sections \ref{section:success-stories} and
\ref{section:failure-of-conjecture}. As a fallout of our studies, we
obtain combinatorial proofs of the classical {\lt} theorem for some of
the aforesaid special cases, and also obtain semantic
characterizations of natural subclasses of the $\Delta^0_2$ fragment
of FO. In Section \ref{section:proof-of-conjecture}, we provide the
proof of the sharper characterization using tools from classical model
theory and some notions that we define. We conclude with questions for
future work in Section \ref{section:conclusion}.

We assume that the reader is familiar with standard notation and
terminology used in the syntax and semantics of FO (see
~\cite{libkin}). A \ti{vocabulary} $\tau$ is a set of predicate,
function and constant symbols. In this paper, we will restrict
ourselves to finite vocabularies only. A \ti{relational vocabulary}
has only predicate and constant symbols, and a \ti{purely relational
  vocabulary} has only predicate symbols.  We denote by $FO(\tau)$,
the set of all FO formulae over vocabulary $\tau$. A sequence $(x_1,
\ldots, x_k)$ of variables is denoted by $\bar{x}$. We will abbreviate
a block of quantifiers of the form $Q x_1 \ldots Q x_k$ by $Q
\bar{x}$, where $Q \in \{\forall, \exists\}$. By $\Sigma^0_k$
(resp. $\Pi^0_k$), we mean FO sentences in Prenex Normal Form (PNF)
over an arbitrary vocabulary, whose quantifier prefix begins with a
$\exists$ (resp. $\forall$) and consists of $k-1$ alternations of
quantifiers.  We use the standard notions of $\tau$-structures,
substructures and extensions, as in ~\cite{libkin}.  Given
$\tau-$structures $M$ and $N$, we denote by $M \subseteq N$ that $M$
is a substructure of $N$ (or $N$ is an extension of $M$). Given $M$
and a subset $S$ (resp. a tuple $\bar{a}$ of elements) of its
universe, we denote by \M{M}{S} (resp. \M{M}{\bar{a}}) the smallest
substructure (under set inclusion ordering of the universe) of $M$
containing $S$ (resp. underlying set of $\bar{a}$) and call it the
substructure of $M$ \emph{induced} by $S$ (resp. underlying set of
$\bar{a}$).  Finally, by \ti{size} of $M$, we mean the cardinality of
its universe and denote it by $|M|$. As a final note of convention,
whenever we talk of FO definability in the paper, we mean definability
via FO sentences (as opposed to theories), unless stated otherwise.

\section{Preservation under substructures modulo  cores}\label{section:pres-under-subst-mod-cores}

We denote by {\PS} the collection of all classes of structures, in any
vocabulary, which are closed under taking substructures. This includes
classes which are not definable in any logic. We let $PS$ denote the
collection of FO definable classes in {\PS}. We identify classes in
$PS$ with their defining FO sentences and will henceforth treat $PS$
as a set of sentences. We now consider a natural generalization of the
{\PS} property. Our discussion will concern arbitrary (finite)
vocabularies and arbitrary structures over them.

\subsection{The case of finite cores}\label{subsection:PSC_f}

\begin{defn}(\tbf{Preservation under substructures modulo finite
  cores})

A class of structures $S$ is said to be \emph{preserved under
  substructures modulo a finite core} (denoted $S \in \PSCf$), if for
every structure $M \in S$, there exists a finite subset $C$ of
elements of $M$ such that if $M_1 \subseteq M$ and $M_1$ contains $C$,
then $M_1 \in S$. The set $C$ is called a \emph{core of $M$
  w.r.t. $S$}. If $S$ is clear from context, we will call $C$ as a
\emph{core of $M$}.
\end{defn}

Note that any finite subset of the universe of $M$ containing a core
is also a core of $M$. Also, there can be multiple cores of $M$ having
the same size. A \ti{minimal} core of $M$ is a core, no subset of
which is a core of $M$.

We will use \PSCf ~to denote the collection of all classes preserved
under substructures modulo a finite core.  Similarly, we will use
$PSC_f$ to denote the collection of FO definable classes in ${\PSCf}$.
We identify classes in $PSC_f$ with their defining FO sentences, and
will henceforth treat $PSC_f$ as a set of sentences.

\vspace*{0.05in}

\tbf{Example 1:} Let $S$ be the class of all graphs containing
cycles. For any graph in $S$, the vertices of any cycle is a core of
the graph. Thus $S \in \PSCf$.

\vspace*{0.05in}

Note that ${\PS} \subseteq {\PSCf}$ since for any class in ${\PS}$ and
for any structure in the class, any element is a core. However it is
easy to check that $S$ in above example is not in ${\PS}$; so
${\PSCf}$ strictly generalizes ${\PS}$. Further, the FO
inexpressibility of $S$ shows that ${\PSCf}$ contains classes not
definable in FO. 

\vspace*{0.05in}

\tbf{Example 2:} Consider $\phi = \exists x \forall y E(x, y)$. In any
graph satisfying $\phi$, any witness for $x$ is a core of the
graph. Thus $\phi \in PSC_f$.  In fact, one can put a uniform bound of
1 on the minimal core size for all models of $\phi$.

\vspace*{0.05in}
Again it is easy to see that $PS \subsetneq PSC_f$.  Specifically, the
sentence $\phi$ in Example 2 is not in $PS$.  This is because a
directed graph with exactly two nodes $a$ and $b$, and having all
directed edges except the self loop on $a$ models $\phi$ but the
subgraph induced by $a$ does not model $\phi$. Hence $PS \subsetneq
PSC_f$.  Extending the example above, one can show that for any
sentence $\varphi$ in $\Sigma^0_2$, in any model of $\varphi$, any
witness for the $\exists$ quantifiers in $\varphi$ forms a core of the
model. Hence $\Sigma^0_2 \subseteq PSC_f$. In fact, for any sentence
in $\Sigma^0_2$, the number of $\exists$ quantifiers serves as a
uniform bound on the minimal core size for all models.  Surprisingly,
even for an arbitrary $\phi \in PSC_f$, it is possible to bound the
minimal core size for all models!

Towards the result, we use the notions of \emph{chain} and \emph{union
  of chain} from the literature. The reader is referred to
~\cite{chang-keisler} for the definitions. We denote a chain as $M_1
\subseteq M_2 \subseteq \ldots$ and its union as $\bigcup_{i \ge 0}
M_i$. We say that a sentence $\phi$ is \ti{preserved under unions of
  chains} if for every chain of models of $\phi$, the union of the
chain is also a model of $\phi$.  We now recall the following
characterization theorem from the '60s ~\cite{chang-keisler}.

\begin{theorem}{(Chang-{\L}o{\'s}-Suszko)~}\label{theorem:PUC}
A sentence $\phi$ is preserved under unions of chains iff it is
equivalent to a $\Pi^0_2$ sentence.
\end{theorem}

Now we have the following theorem.

\begin{theorem}\label{theorem:PSC_f=BSR}
A sentence $\phi \in PSC_f$ iff $\phi$ is equivalent to a $\Sigma^0_2$
sentence.
\end{theorem}
\ti{Proof:} We infer from Theorem \ref{theorem:PUC} the following
equivalences.

$\phi$ is equivalent to a $\Sigma_0^2$ sentence iff

$\neg \phi$ is equivalent to a $\Pi_0^2$ sentence iff

$\forall M_1, M_2, \ldots ((M_1 \subseteq M_2 \subseteq \ldots) \wedge
  (M = \bigcup_{i \geq 1} M_i) \wedge \forall i (M_i \models \neg
  \phi)) \rightarrow M \models \neg \phi$ iff

$\forall M_1, M_2, \ldots ((M_1 \subseteq M_2 \subseteq \ldots) \wedge
  (M = \bigcup_{i \geq 1} M_i) \wedge  (M \models \phi) )
  \rightarrow \exists i (M_i \models \phi)$\\

Assume $\phi \in PSC_f$. Suppose $M_1 \subseteq M_2 \subseteq \ldots$
is a chain, $M = \bigcup_{i \geq 0} M_i$ and $M \models \phi$.  Then,
there exists a finite core $C$ of $M$. For any $a \in C$, there exists
an ordinal $i_a$ s.t. $a \in M_{i_a}$ (else $a$ would not be in the
union $M$). Since $C$ is finite, let $i = \mathbf{max} (i_a |~ a \in
C)$. Since $i_a \leq i$, we have $M_{i_a} \subseteq M_i$; hence $a \in
M_i$ for all $a \in C$.  Thus $M_i$ contains $C$. Since $C$ is a core
of $M$ and $M_i \subseteq M$, $M_i \models \phi$ by definition of
$PSC_f$.  By the equivalences shown above, $\phi$ is equivalent to a
$\Sigma^0_2$ sentence. We have seen earlier that $\Sigma^0_2 \subseteq
PSC_f$.\epf

\begin{Corollary}\label{corollary:finite-core-implies-bounded-core}
If $\phi \in PSC_f$, there exists $B \in \mathbb{N}$ such that every
model of $\phi$ has a core of size atmost $B$.
\end{Corollary}
\ti{Proof}: Take $B$ to be the number of $\exists$ quantifiers in the
equivalent $\Sigma^0_2$ sentence. \epf\\

Given Corollary \ref{corollary:finite-core-implies-bounded-core}, it
is natural to ask if $B$ is computable.  In this context, the
following recent unpublished result by Rossman
~\cite{rossman-los-tarski} is relevant. Let $|\phi|$ denote the size
of $\phi$.
\begin{theorem}(Rossman)\label{theorem:rossman-los-tarski}
There is no recursive function $f: \mathbb{N} \rightarrow \mathbb{N}$
such that if $\phi \in PS$, then there is an equivalent $\Pi^0_1$
sentence of size atmost $f(|\phi|)$. The result holds even for
relational vocabularies and further even if $PS$ is replaced with $PS
\cap \Sigma^0_2$.
\end{theorem}

\begin{Corollary}\label{corollary:los-tarski-no-of-foralls}
There is no recursive function $f: \mathbb{N} \rightarrow \mathbb{N}$
such that if $\phi \in PS$, then there is an equivalent $\Pi^0_1$
sentence with atmost $f(|\phi|)$ universal variables. The result holds
even for relational vocabularies and further even if $PS$ is replaced
with $PS \cap \Sigma^0_2$.
\end{Corollary}
\ti{Proof}: Let $\varphi = \forall^n \bar{z} \psi(\bar{z})$ be a
$\Pi^0_1$ sentence equivalent to $\phi$ where $n = f(|\phi|)$. Let $k$
be the number of atomic formulae in $\psi$. Since $\phi$ and $\psi$
have the same vocabulary, $k \in O(|\phi| \cdot n^{|\phi|})$.  The
size of the Disjunctive Normal Form of $\psi$ is therefore bounded
above by $O(k \cdot n \cdot 2^k)$. Hence $|\varphi|$ is a recursive
function of $|\phi|$ if $f$ is recursive.\epf\\

Theorem \ref{theorem:rossman-los-tarski} strengthens the
non-elementary lower bound given in ~\cite{dawar-model-theory-large}.
Corollary \ref{corollary:los-tarski-no-of-foralls} gives us the
following.

\begin{lemma}
There is no recursive function $f: \mathbb{N} \rightarrow \mathbb{N}$
s.t. if $\phi \in PSC_f$, then every model of $\phi$ has a core of
size atmost $f(|\phi|)$.
\end{lemma}
\ti{Proof}: Consider such a function $f$. For any sentence $\phi$ in a
relational vocabulary $\tau$ s.t. $ \phi \in PS$, $\neg \phi$ is
equivalent to a $\Sigma^0_1$ sentence by {\lt} theorem. Hence $\neg
\phi \in PSC_f$. By assumption about $f$, the size of minimal models
of $\neg \phi$ is bounded above by $n = f(|\phi|) + k$, where $k$ is
the number of constants in $\tau$.  Therefore, $\neg \phi$ is
equivalent to an $\exists^n$ sentence and hence $\phi$ is equivalent
to a $\forall^n$ sentence. Corollary
\ref{corollary:los-tarski-no-of-foralls} now forbids $n$, and hence
$f$, from being recursive. It is easy to see that the result extends
to vocabularies with functions too (by using functions in a trivial
way). \epf\\

Corollary \ref{corollary:finite-core-implies-bounded-core} motivates
us to consider sentences with bounded cores since all sentences in
$PSC_f$ have bounded cores.

\subsection{The case of bounded cores}\label{subsection:PSC(B)-intro}
We first give a more general definition.
\begin{defn}(\tbf{Preservation under substructures modulo a bounded core})
A class of structures $S$ is said to be \emph{preserved under
  substructures modulo a bounded core} (denoted $S \in {\PSC}$), if $S
\in {\PSCf}$ and there exists a finite cardinal $B$ dependent only on
$S$ such that every structure in $S$ has a core of size atmost $B$.
\end{defn}

The collection of all such classes is denoted by ${\PSC}$. Let
${\PSC}(B)$ be the sub-collection of ${\PSC}$ in which each class has
minimal core sizes bounded by $B$. Then ${\PSC} = \bigcup_{B \ge 0}
{\PSC}(B)$. An easy observation is that ${\PSC}(i) \subseteq
{\PSC}(j)$ for $i \leq j$.  As before, ${\PSC}$ and each ${\PSC}(B)$
contain non-FO definable classes. As an example, the class of forests
is in ${\PSC}(0)$. Let $PSC$ (resp. $PSC(B)$) be the FO definable
classes in ${\PSC}$ (resp. ${\PSC}(B)$). \emph{Observe that $PSC(0)$
  is exactly $PS$ and $PSC = \bigcup_{B \ge 0} PSC(B)$}. Therefore,
$PSC$ generalizes $PS$. Further, the hierarchy in $PSC$ is
strict. Consider $\phi \in PSC(k)$ given by $\phi = \exists x_1 \ldots
\exists x_k \bigwedge_{1 \leq i < j \leq k} \neg (x_i = x_j)$. Then
$\phi \notin PSC(l)$ for $l < k$.  From Corollary
\ref{corollary:finite-core-implies-bounded-core}, we have
\begin{lemma}\label{lemma:PSC=PSC_f}
$PSC = PSC_f$.
\end{lemma}

As noted earlier, a $\Sigma^0_2$ sentence $\phi$ with $B$ existential
quantifiers is in $PSC_f$ with minimal core size bounded by $B$. Hence
$\phi \in PSC(B)$. In the converse direction, Theorem
\ref{theorem:PSC_f=BSR} and Lemma \ref{lemma:PSC=PSC_f} together imply
that for a sentence $\phi \in PSC(B)$, there is an equivalent
$\Sigma^0_2$ sentence. We can then ask the following sharper question:
For $\phi \in PSC(B)$, is there an equivalent $\Sigma^0_2$ sentence
having $B$ existential quantifiers?

\begin{theorem}\label{theorem:PSC(B)=E^BA*}
A sentence $\phi \in PSC(B)$ iff it is equivalent to a $\Sigma^0_2$
sentence with $B$ existential quantifiers.
\end{theorem}

The proof of this theorem uses tools from classical model theory and
some notions that we define. We will present it in Section
\ref{section:proof-of-conjecture}.  Before that we shall consider
Theorem \ref{theorem:PSC(B)=E^BA*} for special fragments of FO and for
special classes of structures.  Towards this, we first look at the
notion of relativization from the literature.

\section{Revisiting Relativization}\label{section:relativization}
For purposes of our discussion in this and remaining sections of the
paper, we will assume relational vocabularies (predicates and
constants).

A notion that has proved immensely helpful in proving most of our
positive special cases of Theorem \ref{theorem:PSC(B)=E^BA*} is that
of \emph{relativization}. Informally speaking, given a sentence
$\phi$, we would like to define a formula (with free variables
$\bar{x}$) which asserts that $\phi$ is true in the submodel induced
by $\bar{x}$. The following lemma shows the existence of such a
formula.

\begin{lemma}\label{lemma:finite-relativizations} 
If $\tau$ is a relational vocabulary, for every $FO(\tau)$ sentence
$\phi$ and variables $\bar{x}= (x_1, \ldots, x_k)$, there exists a
{\em quantifier-free} formula $\phi|_{\bar{x}}$ with free variables
$\bar{x}$ such that the following holds: Let $M$ be a model and
$\bar{a}=( a_1, \ldots, a_k)$ be a sequence of elements of $M$. Then

$$(M, a_1, \ldots, a_k) \models \phi|_{\bar{x}} \mbox{~iff~}
\M{M}{\aset} \models \phi$$

\end{lemma}

\ti{Proof}: Let $X = \{x_1, \ldots, x_k\}$ and $C$ be the set of
constants in $\tau$. First replace every $\forall$ quantifier in
$\phi$ by $\neg \exists$. Then replace every subformula of $\phi$ of
the form $\exists x \chi(x, y_1, \ldots, y_k)$ by $\bigvee_{z \in X
  \cup C} \chi(z, y_1, \ldots, y_k)$. \epf\\

We refer to $\phi|_{\bar{x}}$ as `$\phi$ relativized to $\bar{x}$'. We
shall sometimes denote $\phi|_{\bar{x}}$ as $\phi|_{\xset}$ (though
$\bar{x}$ is a sequence and $\xset$ is a set).

We refer to $\phi|_{\bar{x}}$ as `$\phi$ relativized to
$\bar{x}$'. For clarity of exposition, we will abuse notation and use
$\phi|_{\xset}$ to denote $\phi|_{\bar{x}}$ (although $\bar{x}$ is a
sequence and $\xset$ is a set), whenever convenient.

We begin with the following observation.

\begin{lemma}\label{lemma:its-always-possible-to-replace-matrix-by-phi-itself}
Over any given class $\mc{C}$ of structures in \PS, if $\phi
\leftrightarrow \forall z_1 \ldots \forall z_n \varphi$ where
$\varphi$ is quantifier-free, then $\phi \leftrightarrow \psi$ where
$\psi = \forall z_1 \ldots \forall z_n \phi|_{\{z_1, \ldots, z_n\}}$.
\end{lemma}
\ti{Proof}: It is easy to see that $\phi \rightarrow \psi$. Let $M \in
\mc{C}$ be s.t. $M \models \psi$. Let $\bar{a}$ be an $n-$tuple from
$M$. Then, by Lemma~\ref{lemma:finite-relativizations},
$\M{M}{\bar{a}} \models \phi$. Since $\mc{C} \in \PS$, $\M{M}{\bar{a}}
\in \mc{C}$ so that $\M{M}{\bar{a}} \models \forall z_1 \ldots \forall
z_n \varphi$. Then $\M{M}{\bar{a}} \models \varphi(\bar{a})$ and hence
$M \models \varphi(\bar{a})$. Then $M \models \forall z_1 \ldots
\forall z_n \varphi$ and hence $M \models \phi$. \epf\\

Using {\lt} theorem and the above lemma, it follows that a sentence
$\phi$ in $PS$ has an equivalent universal sentence whose matrix is
$\phi$ itself relativized to the universal variables. However we give
a proof of this latter fact directly using relativization, and hence
an alternate proof of the {\lt} theorem. We emphasize that our proof
works only for relational vocabularies ({\lt} is known to hold for
arbitrary vocabularies). This would show that relativization helps us
prove Theorem \ref{theorem:PSC(B)=E^BA*} for the case of $B = 0$.

\subsection{A proof of {\lt} theorem using relativization}\label{subsection:proof-of-los-tarski-using-relativization}

We first introduce some notation. Given a $\tau-$structure $M$, we
denote by $\tau_M$, the vocabulary obtained by expanding $\tau$ with
as many constant symbols as the elements of $M$ - one constant per
element. We denote by $\mc{M}$ the $\tau_M$ structure whose
$\tau-$reduct is $M$ and in which each constant in $\tau_M$ is
interpreted as the element of $M$ corresponding to the constant. It is
clear that $M$ uniquely determines $\mc{M}$. Finally, $\mc{D}(M)$
denotes the \emph{diagram} of $M$ - the collection of quantifier free
$\tau_M-$sentences true in $\mc{M}$.

\begin{theorem}({\lt})\label{theorem:los-tarski-proof-by-bharat-sir}
A FO sentence $\phi$ is in $PS$ iff there exists an $n \in \mathbb{N}$
such that $\phi$ is equivalent to $\forall z_1 \ldots \forall z_n
\phi|_{\{z_1, \ldots, z_n\}}$.
\end{theorem}

\ti{Proof}:

Consider a set of sentences $\Gamma = \{\xi_k \mid k \in \mathbb{N},~
\xi_k = \forall z_1 \ldots \forall z_k \phi|_{\{z_1, \ldots,
  z_k\}}\}$.  Observe that $\xi_{k+1} \rightarrow \xi_k$ so that a
finite collection of $\xi_k$s will be equivalent to $\xi_{k^*}$ where
$k^*$ is the highest index $k$ appearing in the collection. We will
show that $\phi \leftrightarrow \Gamma$. Once we show this, by
compactness theorem, $\phi \leftrightarrow \Gamma_1$ for some finite
subset $\Gamma_1$ of $\Gamma$ and by the preceding observation, $\phi$
is equivalent to $\xi_n \in \Gamma_1$ for some $n$.

If $M \models \phi$, then since $\phi \in PS$, every substructure of
it models $\phi$ - in particular, the substructure induced by any
$k$-elements of $M$. Then $M \models \xi_k$ for every $k$ and hence $M
\models \Gamma$.  

Conversely, suppose $M \models \Gamma$. Then every finite substructure
of $M$ models $\phi$. Let $\mc{M}$ be the $\tau_M$ structure
corresponding to $M$. Consider any finite subset $S$ of the diagram
$\mc{D}(M)$ of $M$. Let $C$ be the finite set of constants referred to
in $S$. Clearly $\mc{M}|_{\tau \cup C}$, namely the $(\tau \cup
C)$-reduct of $\mc{M}$ models $S$ since $\mc{M} \models
\mc{D}(M)$. Then consider the substructure $\mc{M}_1$ of
$\mc{M}|_{\tau \cup C}$ induced by the intepretations of the constants
of $C$ - this satisfies $S$. Now since $C$ is finite, so is
$\mc{M}_1$. Then the $\tau-$reduct of $\mc{M}_1$ - a finite
substructure of $M$ models $\phi$.

Thus $S \cup \{\phi\}$ is satisfiable by $\mc{M}_1$. Since $S$ was
arbitrary, every finite subset of $\mc{D}(M) \cup \{\phi\}$ is
satisfiable so that by compactness, $\mc{D}(M) \cup \{ \phi \}$ is
satisfiable by some structure say $\mc{N}$. Then the $\tau-$reduct $N$
of $\mc{N}$ is s.t. (i) $M$ is embeddable in $N$ and (ii) $N \models
\phi$. Since $\phi \in PS$, the embedding of $M$ in $N$ models $\phi$
and hence $M \models \phi$. \epf\\

The above proof shows that for $\phi \in PS$, there is an equivalent
universal sentence whose matrix is $\phi$ itself, relativised to the
universal variables. In fact, by Lemma
\ref{lemma:its-always-possible-to-replace-matrix-by-phi-itself}, there
is an optimal (in terms of the number of universal variables) such
sentence.

An observation from the proof of Theorem
\ref{theorem:los-tarski-proof-by-bharat-sir} is that, the {\lt}
theorem is true over any class of structures satisfying compactness -
hence in particular the class of structures definable by a FO theory
(indeed this result is known). But there are classes of structures
which are not definable by FO theories but still satisfy compactness:
Consider any FO theory having infinite models and consider the class
of models of this theory whose cardinality is not equal to a given
infinite cardinal. This class satisfies compactness but cannot be
definable by any FO theory due to L\"owenheim-Skolem theorem. Yet
{\lt} theorem would hold over this class.

Having seen the usefulness of relativization in proving Theorem
\ref{theorem:PSC(B)=E^BA*} when $B$ equals 0, it is natural to ask if
this technique works for higher values of $B$ too. We answer this
negatively.

\subsection{Limitations of relativization}\label{subsection:limits-of-relativization}

We show by a concrete example that relativization cannot be used to
prove Theorem \ref{theorem:PSC(B)=E^BA*} in general. This motivates us
to derive necessary and sufficient conditions for relativization to
work.

\tbf{Example 3}: Consider $\phi = \exists x \forall y E(x, y)$ over
$\tau = \{E\}$. Note that $\phi$ is in $PSC(1)$. Suppose $\phi$ is
equivalent to $\psi = \exists x \forall^n \bar{y} \phi|_{x\bar{y}}$
for some $n$. Consider the structure $M = (\mathbb{Z}, \leq)$ namely
the integers with usual $\leq$ linear order. Any finite substructure
of $M$ satisfies $\phi$ since it has a minimum element (under the
linear order). Then taking $x$ to be any integer, we see that $M
\models \psi$. However $M \not\models \phi$ since $M$ has no minimum
element - a contradiction. The same argument can be used to show that
$\phi$ cannot be equivalent to any sentence of the form $\exists^n
\bar{x}~ \forall^m \bar{y} ~ \phi|_{\bar{x}\bar{y}}$. \\

We now give necessary and sufficient conditions for relativization to
work. Towards this, we introduce the following notion.  Consider $\phi
\in FO(\tau)$ s.t. $\phi \in PSC(B)$. Consider a vocabulary $\tau_B$
obtained by expanding $\tau$ with $B$ fresh constants. Consider the
class $S^{\text{all}}_{\phi}$ of $\tau_B$-structures with the
following properties:
\begin{enumerate}
\item For each $(M, a_1, \ldots, a_B) \in S^{\text{all}}_{\phi}$ where
  $M$ is a $\tau-$structure and $a_1, \ldots, a_B \in M$, $M \models
  \phi$ and $\{a_1, \ldots, a_B\}$ forms a core of $M$ w.r.t. $\phi$.
\item For each model $M$ of $\phi$, for each core $C = \{a_1, \ldots,
  a_l\}$ of $M$ w.r.t. $\phi$ s.t.  $l \leq B$ and for each function
  $p: \{1, \ldots, B\} \rightarrow C$ with range $~C$, it must be that
  $(M, p(1), \ldots, p(B)) \in S^{\text{all}}_{\phi}$.
\end{enumerate}

We now have the following.

\begin{theorem}\label{theorem:characterizing-power-of-expansion-with-all-cores-approach}
Given $\phi \in PSC(B)$, the following are equivalent.
\begin{enumerate}
\item $S^{\text{all}}_{\phi}$ is finitely axiomatizable.
\item $\phi$ is equivalent to $\exists^B \bar{x}~ \forall^n \bar{y}
  ~\phi|_{\bar{x}\bar{y}}$ for some $n \in \mathbb{N}$.
\item $\phi$ is equivalent to a $\exists^B \forall^*$ sentence $\psi$
  such that in any model $M$ of $\psi$ and $\phi$, the following hold:
\begin{enumerate}
\item The underlying set of any witness for $\psi$ is a core of $M$
  w.r.t.  $\phi$.
\item Conversely, if $C$ is a core of $M$ w.r.t. $\phi$, $x_1, \ldots,
  x_B$ are the $\exists$ variables of $\psi$ and $f: \{x_1, \ldots,
  x_B\} \rightarrow C$ is any function with range $C$, then $(f(x_1),
  \ldots, f(x_B))$ is witness for $\psi$ in $M$.
\end{enumerate}
\end{enumerate}
\end{theorem}
\ti{Proof}:

\underline{$(1) \rightarrow (2)$}: Let $S^{\text{all}}_{\phi}$ be
finitely axiomatizable. Check that $S^{\text{all}}_{\phi} \in \PS$ so
that by {\lt} theorem, it is axiomatizable by a $\Pi^0_1$
$FO(\tau_B)$-sentence $\psi$ having say $n$ $\forall$
quantifiers. Further, by Lemma
\ref{lemma:its-always-possible-to-replace-matrix-by-phi-itself},
$\psi$ is equivalent to $\gamma = \forall^n \bar{z}
\psi|_{\bar{z}}$. Now consider $\varphi = \exists^B \bar{x}~ \forall^n
\bar{y} ~\phi|_{\bar{x}\bar{y}}$.  Firstly, from Lemma
\ref{lemma:PSC(B)-sentence-implies-relativized-version}, $\phi
\rightarrow \varphi$.  Conversely, suppose $M \models \varphi$. Let
$a_1, \ldots, a_B$ be witnesses and consider the $\tau_B$-structure
$M_B = (M, a_1, \ldots, a_B)$. Now $M_B \models \forall^n \bar{y} ~
\phi|_{\bar{x}\bar{y}}$. We will show that $M_B \models
\gamma$. Consider $b_1, \ldots, b_n \in M$ and let $M_1 =
\M{M_B}{\{b_1, \ldots, b_n\}}$. Then $M_1 \models \forall^n \bar{y}~
\phi|_{\bar{x}\bar{y}}$. Check that the $\tau-$reduct of $M_1$ (i)
models $\phi$ and (ii) contains $\{a_1, \ldots, a_B\}$ as a core. Then
$M_1 \in S^{\text{all}}_{\phi}$ and hence $M_1 \models \psi$. Since
$b_1, \ldots, b_n$ were arbitrary, $M_B \models \gamma$. Since $\gamma
\leftrightarrow \psi$ and $\psi$ axiomatizes $S^{\text{all}}_{\phi}$,
the $\tau-$reduct of $M_B$, namely $M$, models $\phi$.\\

\underline{$(2) \rightarrow (3)$}: Take $\psi$ to be $ \exists^B
\bar{x}~ \forall^n \bar{y}~ \phi|_{\bar{x}\bar{y}}$. Consider a model
$M$ of $\phi$ and $\psi$.  The set $C$ of elements of any witness for
$\psi$ forms a core of $M$ w.r.t. $\psi$. Then since $\phi
\leftrightarrow \psi$, $C$ is also a core of $M$
w.r.t. $\phi$. Conversely, consider a core $C$ of $M$ w.r.t.
$\phi$. Then any substructure of $M$ containing $C$ satisfies
$\phi$. Then check that elements of $C$ form a witness for $\psi$.\\

\underline{$(3) \rightarrow (1)$}: Let $\phi \leftrightarrow \psi$
where $\psi = \exists^B \bar{x}~ \forall^n \bar{y} \beta (\bar{x},
\bar{y})$ where $\beta$ is quantifier free and $\psi$ satisfies the
conditions mentioned in (3). Consider $\varphi = \forall^n \bar{y}
~\beta[x_1 \mapsto c_1, \ldots, x_B \mapsto c_B]$ where $c_1, \ldots,
c_B$ are $B$ fresh constants and $x_i \mapsto c_i$ means replacement
of $x_i$ by $c_i$. If $M_B = (M, a_1, \ldots, a_B) \models \varphi$,
then $M \models \psi$ and hence $M \models \phi$. Since $a_1, \ldots,
a_B$ are witnesses for $\psi$ in $M$, they form a core of $M$
w.r.t. $\phi$ by assumption, so that $M_B \in S^{\text{all}}_{\phi}$.
Conversely, if $M_B = (M, a_1, \ldots, a_B) \in
S^{\text{all}}_{\phi}$, then $M \models \phi$ and $a_1, \ldots, a_B$
form a core in $M$. Then by assumption, $M \models \psi$ and $a_1,
\ldots, a_B$ are witnesses for $\psi$. Then $M_B \models \varphi$. To
sum up, $\varphi$ axiomatizes $S^{\text{all}}_{\phi}$.\epf\\

Consider $\phi$ and $M$ in the Example 3 above. Take any finite
substructure $M_1$ of $M$ - it models $\phi$. There is exactly one
witness for $\phi$ in $M_1$, namely the least element under
$\leq$. However every element in $M_1$ serves as a core. The above
theorem shows that no $\exists \forall^*$ sentence will be able to
capture exactly all the cores through its $\exists$ variable.

In the following sections, we shall study Theorem
\ref{theorem:PSC(B)=E^BA*} for several special classes of FO and over
special structures. Interestingly, in most of the cases in which
Theorem \ref{theorem:PSC(B)=E^BA*} turns out true, relativization
works! However we also show a case in which relativization does not
work, yet Theorem \ref{theorem:PSC(B)=E^BA*} is true.

\section{Positive Special Cases for Theorem \ref{theorem:PSC(B)=E^BA*}}\label{section:success-stories}
\subsection{Theorem \ref{theorem:PSC(B)=E^BA*} holds for special fragments of FO}\label{subsection:PSC(B)-special-fragments}
Unless otherwise stated, we consider relational vocabularies
throughout the section.  The following lemma will be repeatedly used
in the subsequent results.

\begin{lemma}\label{lemma:PSC(B)-sentence-implies-relativized-version}
Let $\phi \in PSC(B)$. For every $n \in \mathbb{N}$, $\phi$ implies
$\exists^B \bar{x} ~\forall^n \bar{y} ~\phi|_{\bar{x}\bar{y}}$.
\end{lemma}
\ti{Proof}: Suppose $M \models \phi$.  Since $\phi \in PSC(B)$, there
is a core $C$ of $M$ of size at most $B$. Interpret $\bar{x}$ to
include all the elements of $C$ (in any which way). Since $C$ is a
core, for any $n$-tuple $\bar{d}$ of elements of $M$, having
underlying set $D$, the substructure of $M$ induced by $C \cup D$
models $\phi$. Then $(M, \bar{a}, \bar{d}) \models
\phi|_{\bar{x}\bar{y}}$ for all $\bar{d}$ from $M$. \epf\\

\begin{lemma}\label{lemma:conjecture-for-monadic-case}
Let $\tau$ be a monadic vocabulary containing $k$ unary
predicates. Let $\phi \in FO(\tau)$ be a sentence of rank $r$
s.t. $\phi \in PSC(B)$. Then $\phi$ is equivalent to $\psi$ where
$\psi = \exists^B \bar{x} ~\forall^n \bar{y}~ \phi|_{\bar{x}\bar{y}}$
where $n = r \times 2^k$. For $B = 0$, $n$ is optimal i.e. there is an
FO sentence in $PSC(0)$ for which any equivalent $\Pi^0_2$ sentence
has atleast $n$ quantifiers.
\end{lemma}
\ti{Proof}: That $\phi$ implies $\psi$ follows from Lemma
\ref{lemma:PSC(B)-sentence-implies-relativized-version}. For the
converse, suppose $M \models \psi$ where $n = r \times 2^k$. By an
Ehrenfeucht-Fr\"aiss\'e game argument, we can show that $M$ contains a
substructure $M_S$ such that (i) $M \equiv_r M_S$, with $|M_S| \leq n$
and (ii) for any extension $M'$ of $M_S$ in $M$, $M' \equiv_r
M_S$. The substructure $M_S$ is obtained by taking up to $r$ elements
of each colour $c \in 2^\tau$ present in $M$. An element $a$ in
structure $M$ is said to have colour $c$ if for every predicate $P \in
\Sigma$, $M \models P(a)$ iff $P \in c$.  Since $M \models \psi$,
there exists witnesses $\bar{a}$ for $\psi$ in $M$. Choose $\bar{b}$
to be an $n$-tuple which includes the elements of $M_S$. This is
possible because $|M_S| \le n$. Then we have, $(M, \bar{a}, \bar{b})
\models \phi|_{\bar{x}\bar{y}}$ so that $\M{M}{\bar{a}\bar{b}} \models
\phi$.  But $M_S \subseteq \M{M}{\bar{a}\bar{b}} \subseteq M$ so that
$\M{M}{\bar{a}\bar{b}} \equiv_r M$. Then $M \models \phi$.\\

To see the optimality of $n$ for $B = 0$, consider the sentence $\phi$
which states that there exists at least one colour $c \in 2^\tau$ such
that there exist at most $r-1$ elements with colour $c$. The sentence
$\phi$ can be written as a formula with rank $r$, as the disjunction
over all colours, of sentences of the form, $\exists x_1 \exists x_2
\cdots \exists x_{r-1} \forall x_r (\bigwedge_{i=1}^{r-1} x_r \ne x_i)
\rightarrow \lnot C(x_r)$.  From the preceding paragraph, $\phi
\leftrightarrow \forall^n \bar{y} ~ \phi|_{\bar{y}}$ where $n = r
\times 2^k$. Suppose $\phi$ is equivalent to a $\forall^s$ sentence
for some $s < n$. Then by Lemma
\ref{lemma:its-always-possible-to-replace-matrix-by-phi-itself}, $\phi
\leftrightarrow \varphi$ where $\varphi = \forall^{s} \bar{y} ~
\phi|_{\bar{y}}$. Then consider the structure $M$, which has $r$
elements of each colour. Clearly, $M \not\models \phi$. However check
that every $s$-sized substructure of $M$ models $\phi$. Then $M
\models \varphi$ and hence $M \models \phi$ - a contradiction.\epf

\begin{lemma}\label{lemma:conjecture-for-a-finite-set-of-finite-structures}
Let $S \in {\PSC}(B)$ be a finite collection of $\tau-$structures so
that $S$ is definable by a $\Sigma^0_2$ sentence $\phi \in
PSC(B)$. Then $S$ is definable by the sentence $\psi$ where $\psi =
\exists^B \bar{x} ~\forall^n \bar{y}~ \phi|_{\bar{x}\bar{y}}$ for some
$n \in \mathbb{N}$.
\end{lemma}

\ti{Proof}: Check that all structures in $S$ must be of finite size so
that $\phi$ exists. Let the size of the largest structure in $S$ be
atmost $n$. Consider $\psi$.  Lemma
\ref{lemma:PSC(B)-sentence-implies-relativized-version} shows that
$\phi \rightarrow \psi$. Conversely, suppose $M \models \psi$. Then
there exists a witness $\bar{a}$ s.t. any extension of
$\M{M}{\bar{a}}$ within $M$ with atmost $n$ additional elements models
$\phi$. Since $M$ is of size atmost $n$, taking the extension $M$ of
$\M{M}{\bar{a}}$, we have $M \models \phi$. Since $\phi$ defines $S$
so does $\psi$.\epf

\begin{lemma}\label{lemma:conjecture-for-Pi^0_2} 
Consider $\phi \in \Pi^0_2$ given by $\phi = \forall^n \bar{x} ~
\exists^m \bar{y} ~\beta (\bar{x}, \bar{y})$ where $\beta$ is
quantifier free. If $\phi \in PSC(B)$, then $\phi$ is equivalent to
$\psi$ where $\psi = \exists^B \bar{u} ~\forall^n \bar{v}~
\phi|_{\bar{u}\bar{v}}$.
\end{lemma}

\ti{Proof}: From Lemma
\ref{lemma:PSC(B)-sentence-implies-relativized-version}, $\phi
\rightarrow \psi$. For the converse, let $M \models \psi$ and let
$\bar{a}$ be a witness. Consider an $n-$tuple $\bar{b}$ from $
M$. Then $M_1 = \M{M}{\bar{a}\bar{b}}$ is s.t. $M_1 \models
\phi$. Then for $\bar{x} = \bar{b}$, there exists $\bar{y} = \bar{d}$
s.t. $\bar{d}$ is an $m-$tuple from $ M_1$ and $M_1 \models
\beta(\bar{b}, \bar{d})$. Then $M \models \beta(\bar{b}, \bar{d})$
since $M_1 \subseteq M$. Hence $M \models \phi$.\epf

\begin{lemma}\label{lemma:conjecture-when-phi-and-neg-phi-are-in-PSC}
Suppose $\phi \in PSC(B)$ and $\neg \phi \in PSC(B')$. Then $\phi$ is
equivalent to $\psi$ where $\psi = \exists^B \bar{x} ~\forall^{B'}
\bar{y}~ \phi|_{\bar{x}\bar{y}} $.
\end{lemma}
\ti{Proof}: From Lemma
\ref{lemma:PSC(B)-sentence-implies-relativized-version}, $\phi$
implies $\psi$. For the converse, suppose $M \models \psi$. Then there
is a witness $\bar{a}$ for $\psi$ s.t. for any $B'$-tuple $\bar{b}$,
the substructure induced by $\bar{a}\bar{b}$
i.e. $\M{M}{\bar{a}\bar{b}}$ models $\phi$. Suppose $M \not\models
\phi$. Then $M \models \neg \phi$ so that there is a core $C$ of $M$
w.r.t. $\neg \phi$, of size at most $B'$. Let $\bar{d}$ be a $B'$-tuple
which includes all the elements of $C$. Then $\M{M}{\bar{a}\bar{d}}
\models \phi$. But $\M{M}{\bar{a}\bar{d}} \subseteq M$ contains $C$ so
that $\M{M}{\bar{a}\bar{d}} \models \neg \phi$ -- a
contradiction.\epf\\

Observe that for the special case of $B = 0$, we get combinatorial
proofs of {\lt} theorem for the fragments mentioned above. Moreover
all of these proofs and hence the results hold in the finite.  We
mention that the result of Lemma \ref{lemma:conjecture-for-Pi^0_2}
holding in the finite was proved by Compton too (see
~\cite{gurevich-shelah}). We were unware of this until recently and
have independently arrived at the same result. The reader is referred
to Section \ref{section:los-tarski-additional-observations} for our
studies on more \emph{positive} cases of {\lt} in the finite.

Interestingly, Lemma
\ref{lemma:conjecture-when-phi-and-neg-phi-are-in-PSC} has
implications for the $\Delta^0_2$ fragment of FO. Define
$\Delta^0_2(k, l) \subseteq \Delta^0_2$ to be the class of sentences
which have a $\exists^k \forall^*$ and a $\forall^l \exists^*$
equivalent. Note that $\Delta^0_2 = \bigcup_{l, k \geq 0}
\Delta^0_2(k, l)$. Lemma
\ref{lemma:conjecture-when-phi-and-neg-phi-are-in-PSC} gives us the
following right away.

\begin{theorem}\label{theorem:Delta^0_2}
The following are equivalent:
\begin{enumerate}
\item $\phi \in PSC(k)$ and $\neg \phi \in PSC(l)$.
\item $\phi$ is equivalent to a $\exists^k \forall^l$ and a $\forall^l
  \exists^k$ sentence.
\item $\phi \in \Delta^0_2(k, l)$.
\end{enumerate}
\end{theorem}

As a corollary, we see that $\Delta^0_2(k, l)$ is a finite class upto
equivalence. We are not aware of any other semantic characterization
of these natural fragments of $\Delta^0_2$. This highlights the
importance of the notion of cores and the sizes thereof.

\subsection{Theorem \ref{theorem:PSC(B)=E^BA*} over special classes of structures}\label{subsection:PSC(B)-special-classes-of-structures}

We first look at Theorem \ref{theorem:PSC(B)=E^BA*} over finite words
which are finite structures in the vocabulary containing one binary
predicate $\leq$ (always interpreted as a linear order) and a finite
number of unary predicates (which form a partition of the
universe). And we obtain something stronger than Theorem
\ref{theorem:PSC(B)=E^BA*}. Before that, we mention that the idea of
relativization can be naturally extended to MSO. Given $\phi$ in MSO
and a set of variables $Z = \{z_1, \ldots, z_n\}$, $\phi|_{Z}$ is
obtained by first converting all $\forall X$ to $\neg \exists X$ and
then replacing every subformula $\exists X \chi(X, \ldots)$ with
$\bigvee_{Y \subseteq Z} ((\bigwedge_{z \in Y} X(z) \wedge
\bigwedge_{z \in Z \setminus Y} \neg X(z) ) \wedge \chi(X,
\ldots))$. The resulting \emph{FO} formula is then relativized to $Z$
and simplified to eliminate the (original) SO variables. As before,
abusing notation, we use $\phi|_Z$ and $\phi|_{\bar{z}}$
interchangeably.

Note: We at times will refer to the `structure' connotation of a word
and at other times refer to the `string' connotation of it. This would
however be clear from the context (typically language-theoretic
notions used for a word would mean we are talking about it as a string
whereas model-theoretic notions used for it would mean we are
referring to it as a structure).

\begin{theorem}\label{theorem:conjecture-over-words-the-automata-way}
Over words, a MSO sentence $\phi$ is in ${\PSC}(B)$ iff it is
equivalent to $\psi$ where $\psi = \exists^B \bar{x} \forall^k \bar{y}
\phi|_{\bar{x}\bar{y}}$ for some $k \in \mathbb{N}$.
\end{theorem}
\ti{Proof sketch}: We use the fact that over words, by the
B\"uchi-Elgot-Trakhtenbrot theorem ~\cite{buchi}, $MSO$ sentences
define regular languages. The `If' direction is easy. For the `Only
if' direction, let the regular language $L$ defined by $\phi$ be
recognized by an $n$ state automaton, say $\mathcal{M}$.  If there is
no word of length $> N = (B+1) \times n$ in $L$, then $L$ is a finite
language of finite words and hence from Lemma
\ref{lemma:conjecture-for-a-finite-set-of-finite-structures}, we are
done.  Else suppose there is a word of length $> N$ in $L$. Then
consider $\psi$ above for $k = N$.  It is easy to observe that $\phi$
implies $\psi$. In the other direction, suppose $w \models \psi$ for
some word $w$. Then there exists a set $A$ of elements $i_1, \ldots,
i_m$ s.t. (i) $m \leq B$ and $i_1 < i_2 \cdots < i_m$ and (ii) every
substructure of $w$ of size atmost $N+m$ containing $A$ models
$\phi$. From Lemma
\ref{lemma:key_lemma_for_conjecture_for_regular_languages} below,
there exists a substructure $w_1$ of $w$ containing $A$ such that (i)
$|w_1| \leq N$ and (ii) $w_1 \in L$ iff $w \in L$.  Then $w_1$ models
$\phi$ and hence $w \models \phi$. Thus $\psi$ implies $\phi$ and
hence is equivalent to $\phi$.\epf\\

Before going into the proof of the lemma, we briefly explain the
intuition. Let $q_j$ be the state reached by automaton $\mc{M}$ upon
reading the subword $w[1\ldots i_j]$. The subword $w[(i_j+1), \ldots
  i_{j+1}]$ takes $\mc{M}$ from $q_j$ to $q_{j+1}$ through a sequence
$S$ of states. Since $\mc{M}$ has only $n$ states, if $w[(i_j+1),
  \ldots i_{j+1}]$ is long, then $S$ will contain at least one
loop. Then getting rid of the subwords that give rise to loops, we
will be able to obtain a subword of $w[(i_j+1), \ldots i_{j+1}]$ that
takes $\mc{M}$ from $q_j$ to $q_{j+1}$ without causing $\mc{M}$ to
loop in between. It follows that this subword must be of length at
most $n$. Collecting such subwords of $w[(i_j+1), \ldots i_{j+1}]$ for
each $j$ and concatenating them, we get a subword of $w$ of length at
most $N$ containing set $A$ that takes $\mc{M}$ from the initial state
to the same state as $w$.  We now formalize this intuition.

\begin{lemma}\label{lemma:key_lemma_for_conjecture_for_regular_languages}
Let $L$ be a regular language having an $n$ state automaton accepting
it. Given a natural number $B$, consider a word $w \in \Sigma^*$ of
length $> N = (B+1) \times n$.  Let $A = \{i_1, \ldots, i_m\}$ where
$i_1 < i_2 \ldots < i_m$ be a given set of elements from the universe
of $w$. Then there is a substructure $w_1$ of $w$ containing $A$ such
that (i) $|w_1| \leq N$ and (ii) $w_1 \in L$ iff $w \in L$.
\end{lemma}

\ti{Proof}:

Let $M = (Q, \Sigma, \delta, q_0, F)$ be a DFA accepting $L$ where $Q
= \{q_0, \ldots, q_{n-1}\}$ is the set of states, $\Sigma$ is the
alphabet, $\delta$ is the transition function, $q_0$ is the initial
state and $F$ is the set of final states. We use the following
notation: If $z$ is a sequence of objects, then we use $z(k)$ to
denote the $k^{th}$ element of $z$ and $z\left[k \ldots l\right]$ to
denote the subsequence of $z$ formed by the $k^{th}, (k+1)^{th},
\ldots l^{th}$ elements of $z$ for $k, l$ s.t. $1 \leq k \leq l \leq $
(length of $z$).

Let $q(i+1),~ 1 \leq i \leq |w|$ be the state of $Q$ after reading the
word $w\left[1\ldots i\right]$. We take $q(1)$ to be $q_0$. Then let
$q = (q(i))_{1 \leq i \leq (|w|+1)}$ be the sequence of these states.
We are given $A = \{i_1, \ldots, i_m\}$ which is a subset of $m$
elements of the universe of $w$.  Let $i_0 = 1$ and $i_{m+1} =
|w|+1$. For $j \in \{0, \ldots, m\}$, consider $q\left[i_j \ldots
  i_{j+1}\right]$. Set $p = i_j$ to $s = i_{j+1} - 1$. We collect a
set $T$ of indices between $p$ and $s$ using the procedure below:

Initialize $i$ to $p$.

\begin{enumerate}
\item If $i > s$, then stop.
\item If $i = s$, then put $i$ into $T$ and increment $i$ by 1.
\item If $i < s$, then let $k$ s.t. $p \leq k \leq s$ be the highest
  index such that $q(i) = q(k)$. Then put $k$ into $T$ and update the
  value of $i$ to be $k+1$.
\end{enumerate}

At the end of this procedure, let the indices in $T$ be $k_1, \ldots,
k_l$ where $k_1 < k_2 < \cdots < k_l$ if $T$ is non-empty. Note that
$T$ is empty iff $i_j = i_{j+1}$ only if $j = 0$. Also note that at
termination, the value of $i$ must be $s+1$. Finally note that
$q(i_j), q(k_1+1), q(k_2+1), \ldots, q(k_l)$ must all be distinct so
that $l \leq n$.

Then consider the subword $w_j$ of $w$ given by

\begin{equation}
w_j = \begin{cases}
\epsilon & \text{if T is empty}\\
w(k_1) \cdot w(k_2) \cdots w(k_l) & \text{if T is non-empty}\\ 
\end{cases}\nonumber
\end{equation}

Observe that $|w_j| \leq n$. Let $r_1, \ldots, r_l$ be the states the
automaton $M$ goes through when $w_j$ is applied to state $q(i_j)$.

We consider the following cases:
\begin{enumerate}
\item $T$ is non-empty. 

  Now from the way $k_1$ was chosen by the above procedure, $q(i_j) =
  q(k_1)$. Then if $M$ is in state $q(i_j)$, on $w(k_1)$, it moves to
  state $r_1$ given by $r_1 = q(k_1 + 1)$. Similarly, the index $k_2$
  is s.t. $q(k_2) = q(k_1 +1)$ so that if $M$ is in state $r_1$, then
  on $w(k_2)$, it moves to state $r_2$ given by $r_2 = q(k_2 +
  1)$. Continuing this way we find that on $w(k_l)$, if $M$ is in
  state $r_{l-1}$, it moves to state $r_l$ given by $r_l = q(k_l +
  1)$. Now as observed above, at termination, the value of $i$ must be
  $s + 1 = i_{j+1}$. This can happen in only two ways: (a) In the
  previous iteration of the procedure, step (2) was executed in which
  case $s$ was put in $T$ - then $k_l = s$. (b) In the previous
  iteration of the procedure, step (3) was executed in which case $s$
  again was put into $T$ so that $k_l = s$. Then in either case $k_l =
  s = i_{j+1} - 1$ so that $r_l = q(i_{j+1})$.

  Thus we see that both $w_j$ and $w\left[i_j \ldots (i_{j+1} -
    1)\right]$, when applied to $M$ in state $q(i_j)$, take $M$ to the
  same state, namely $q(i_{j+1})$.

\item $T$ is empty. 

  Then $w_j = \epsilon$ and $i_j = i_{j+1}$ in which case $w\left[i_j
    \ldots (i_{j+1} - 1)\right] = \epsilon$ so that both these words
  applied to $M$ in state $q(i_j)$, take $M$ to the same state, namely
  $q(i_{j+1})$.
\end{enumerate}

Then consider the word $w_1 = w_0 \cdot w_1 \cdots w_m$. From the
above observations, it follows that $w_1$ applied to the initial state
of $M$ takes $M$ to the same state as $w$. Then $w_1 \in L$ iff $w \in
L$. Further since for each $j, ~ |w_j| \leq n$, we have that $|w_1|
\leq (m + 1) \times n \leq (B + 1) \times n = N$. \epf\\

Returning to Theorem
\ref{theorem:conjecture-over-words-the-automata-way}, observe that for
the special case of $B = 0$, we obtain {\lt} theorem for words and
also give a bound for the number of $\forall$s in the equivalent
$\Pi^0_1$ sentence in terms of the number of states of the automaton
for $\phi$. We have not encountered this result in our literature
survey.

Before proceeding ahead, as a slight diversion, we give a simpler
proof of {\lt} theorem over words. In fact, over words, we have the
following stronger result.

\begin{lemma}\label{lemma:los-tarski-over-words-the-higman-way}
Consider any set $S$ of words which is closed under taking
substructures. Then $S$ can be defined by a $\Pi^0_1$ sentence.
\end{lemma}

\ti{Proof}: Consider $\overline{S} = \Sigma^* \setminus S$ - the
complement of $S$. Since $S$ is closed under taking substructures,
$\overline{S}$ is closed under taking extensions.  Then consider the
set $T$ of minimal words of $\overline{S}$, i.e. words of
$\overline{S}$ for which no subword is contained in $\overline{S}$. We
show that $T$ must be finite. Suppose $T$ were infinite.  If we
arrange the words of $T$ to form a sequence - which is infinite - then
by Higman's lemma, there is some word in the sequence which is a
subword of another in the sequence. That means some word of $T$ is a
subword of another word in $T$. But that contradicts the minimality of
the latter word.

Then $T$ is finite. Taking the disjunction of the existential closures
of the diagrams of the words of $T$, we get a $\Sigma^0_1$ sentence
defining $\overline{S}$. Then taking the negation of this sentence, we
get the desired $\Pi^0_1$ sentence defining $S$. \epf\\

Thus contrary to the general setting where it is not necessary for a
set of structures preserved under substructures to be even
FO-expressible, leave alone being definable by a $\Pi^0_1$ sentence,
over words, $\Pi^0_1$ sentences show much greater power.\\

We return to Theorem \ref{theorem:PSC(B)=E^BA*} now. So far,
relativization has worked in all the cases we have seen. We now give
an example of a class of structures over which relativization fails,
yet Theorem \ref{theorem:PSC(B)=E^BA*} is true.

Consider a subclass $\mc{C}$ of bounded degree graphs in which each
graph is a collection (finite or infinite) of \emph{oriented} paths
(finite or infinite). For clarity, by oriented path we mean a graph
isomorphic to a connected induced subgraph of the graph $(V, E)$ where
$V = \mathbb{Z}$ and $E = \{(i, i+1)\,| \,i \in \mathbb{Z}\}$. Observe
that $\mc{C}$ can be axiomatized by a theory $\mc{T}$ which asserts
that every node has in-degree atmost 1 and out-degree atmost 1 and
that there is no directed cycle of length $k$ for each $k \ge 0$. We
first show the following.

\begin{lemma}\label{lemma:failure-of-relativization-over-directed-paths}
For each $B \ge 1$, there is a sentence $\phi \in PSC(B)$ which is not
equivalent, over $\mc{C}$, to any $\psi$ of the form $\exists^B
\bar{x} ~\forall^n \bar{y}~ \phi|_{\bar{x}\bar{y}}$ .
\end{lemma}
\ti{Proof}: Consider $\phi$ which asserts that there are atleast $B$
elements of \emph{total} degree atmost 1 where total degree is the sum
of in-degree and out-degree. Clearly $\phi \in PSC(B)$ since it is
expressible as a $\exists^B \forall^*$ sentence. Suppose $\phi$ is
equivalent to $\psi$ of the form above for some $n \in
\mathbb{N}$. Consider $M \in \mc{C}$ which is a both-ways infinite
path so that every node in $M$ has total degree 2 - then $M
\not\models \phi$. Consider $B$ distinct points on this path at a
distance of atleast $2n$ from each other and form a $B-$tuple say
$\bar{a}$ with them. Let $\bar{b}$ be any $n-$tuple from $M$. Now
observe that $\M{M}{\bar{a}\bar{b}}$ is a finite structure which has
atleast $B$ distinct paths (0-sized paths included). Then
$\M{M}{\bar{a}\bar{b}} \models \phi$ so that $(M, \bar{a}, \bar{b})
\models \phi|_{\bar{x}\bar{y}}$. Since $\bar{b}$ was arbitrary, $M
\models \psi$ so that $M \models \phi$. Contradiction. \epf\\

However Theorem \ref{theorem:PSC(B)=E^BA*} holds over $\mc{C}$!  

\begin{theorem}\label{theorem:conjecture-holds-over-directed-paths}
Over the class $\mc{C}$ of graphs defined above, $\phi \in PSC(B)$ iff
$\phi$ is equivalent to a $\exists^B \forall^*$ sentence.
\end{theorem}
\ti{Proof}: If $\tau = \{E\}$ is the vocabulary of $\phi$, let
$\tau_B$ be a vocabulary obtained by addding $B$ fresh constants $c_1,
\ldots, c_B$ to $\tau$. Given a class $\mc{S}$ of $\tau-$structures,
define $\mc{S}_B$ to be the class of \emph{all} $\tau_B-$structures
s.t. the $\tau-$reduct of each structure in $\mc{S}_B$ is in $\mc{S}$.
Then the proof can be divided into two main steps. Below $\equiv$
denotes elementary equivalence.

\textbf{Step 1} : Given $\phi$, define class $\mathcal{C}' \subseteq
\mathcal{C}$ such that for every structure $A \in \mathcal{C}_B$,
there exists structure $D \in \mathcal{C}'_B$ such that $A \equiv D$
(Property I). Since compactness theorem holds over $\mathcal{C}_B$ (as
$\mc{C}_B$ is defined by the same theory $\mc{T}$ as $\mc{C}$), it
also holds over $\mathcal{C}'_B$.
 
\textbf{Step 2} : Show that $\phi$ is equivalent to an $ \exists^B
\forall^*$ sentence over $\mathcal{C}'$, hence showing the same over
$\mathcal{C}$ as well.

Note: The conditions in \tbf{Step 1} imply that for every $A \in
\mathcal{C}$, there exists $D \in \mathcal{C}'$ such that $A \equiv
D$. Then since compactness theorem holds over $\mathcal{C}$, it also
holds over $\mathcal{C}'$.

\noindent Suppose the rank of $\phi$ is $m$. We define $\mathcal{C}'$
to be the set of graphs $G \in \mathcal{C}$ such that either (a) there
exists a bound $n_G$ (dependent on $G$) such that all paths in $G$
have length less than $n_G$ (this does not mean that $G$ is finite -
there could be infinite paths of the same length in $G$) or (b) there
are atleast $(B + m + 2)$ paths in $G$ which are infinite in both
directions.  It can be shown that $\mathcal{C'}$ satisfies Property I
(See \tbf{A} below). We proceed assuming this to be true.
  
\noindent Now, to show \tbf{Step 2}, we use the following approach.

Let $P \in \mc{C}'$ be s.t. $P \models \phi$. Choose a core $Z$ in $P$
(recall that $\phi \in PSC(B)$). Let $M_P \in \mc{C}'_B$ be a
$\tau_B-$structure whose $\tau-$reduct is $P$ and in which each
element of $Z$ is assigned to some constant. Let $\Gamma^{M_P}$ be the
set of all $\forall^*$ sentences true in $M_P$.  We can show that if
$M' \in \mathcal{C}'_B$ is such that $M' \models \Gamma^{M_P}$, then
$M' \models \phi$ (See \tbf{B} below. We proceed assuming this to be
true). That is, if every finite substructure of $M'$ is embeddable in
$M_P$, then $M' \models \phi$. Then over $\mc{C}'_B$, $\Gamma^{M_P}
\rightarrow \phi$. Now, since $\mathcal{C}'_B$ satisfies compactness
theorem, there exists a finite subset $\Gamma_0^{M_P}$ of
$\Gamma^{M_P}$ such that $\Gamma_0^{M_P} \rightarrow \phi$ over
$\mc{C}'_B$. Note that, since $\Gamma_0^{M_P}$ is a conjunction of
$\forall^*$ sentences, we can assume that $\Gamma_0^{M_P}$ is a single
$\forall^*$ sentence.

Let $\phi_P$ be the $\tau-$sentence of the form $\exists^B \forall^*$
obtained by replacing the $B$ constants in $\Gamma_0^{M_P}$ by $B$
fresh variables and existentially quantifying these variables. Then
check that $\phi_P \rightarrow \phi$.  It is easy to see that $\phi
\rightarrow \bigvee_{P \in \mathcal{C}', P \models \phi} \phi_P$ (If
$P \models \phi$, then interpret the $\exists$ quantifiers in $\phi_P$
as the chosen core $Z$ mentioned above). By compactness theorem over
$\mathcal{C}'$, there exists a finite set of structures, say $\{P_1,
\cdots, P_m\}$ such that $P_i \in \mathcal{C}'$, $P_i \models \phi$
and $\phi \rightarrow \bigvee_{i=1}^{i=m} \phi_{P_i}$. Then, we have
$\phi \leftrightarrow \bigvee_{i=0}^{i=m} \phi_{P_i}$ over
$\mathcal{C}'$. Since each $\phi_{P_i}$ is of the form
$\exists^B\forall^*$, $\bigvee_{i=0}^{i=m} \phi_{P_i}$ is also of the
same form. That completes \textbf{Step 2} and completes the proof.\\

Below we shall be referring to the notions of `ball type of radius
$r$' (or simply $r-$ball type), `disjoint unions' (denoted by
$\sqcup$) and `$m$-equivalence' (denoted by $\equiv_m$). We shall also
use Hanf's theorem. The reader is referred to ~\cite{libkin} for these
concepts.\\

\underline{\tbf{A}. ~$\mathcal{C'}$ satisfies Property I}

\noindent Suppose $A \in \mathcal{C}_B$. If there exists a bound
$n_A$, such that all paths in $A$ have length less than $n_A$, then $A
\in \mathcal{C}'_B$ and hence we are done. Contrarily, suppose that
there is no such bound $n_A$. This means that either there are paths
of arbitrarily large lengths in $A$ or there is atleast one infinite
path in $A$ (Let us mark this inference as [*]). Now, construct
structure $D \in \mathcal{C}'_B$, where $D = A \sqcup
\bigsqcup_{i=1}^{k+m+2} P$, where $P$ is a path which is infinite in
both directions and $\sqcup$ denotes disjoint union. We show that $A
\equiv D$, by showing that for every $n \in \mathbb{N}$, $A \equiv_n
D$.  By Hanf's theorem ~\cite{libkin}, given $n$, there exist numbers
$r$ and $q$, dependent only on $n$, such that $A \equiv_n D$ if for
each ball type $\xi$ of radius $r$, the number of instances of $\xi$
in $A$ and $D$ are either equal or are both are greater than $q$. By
adding $(B + m + 2)$ paths, we are introducing infinite copies of just
one $r-$ball type $\xi$ in $D$, namely the $2r + 1$ length path with
the ball center as the midpoint. However, this type $\xi$ was already
present infinitely many times in $A$ (due to [*]). Hence Hanf's
condition holds for every type $\xi$, and thus, $A \equiv D$.\\

\underline{\tbf{B}. ~If $M_1 \in \mathcal{C}'_B$ is such that $M_1
  \models \Gamma^{M_P}$, then $M_1 \models \phi$}

\noindent Before, we proceed, we state and prove the following
lemma. Below, an `outwardly' (resp. `inwardly') infinite path is an
oriented infinite path with an end point which has an outgoing
(resp. incoming) edge and no incoming (resp. outgoing) edge.
\begin{lemma}\label{lemma:m-sig}
    For every $m \in \mathbb{N}$ and structure $G \in \mathcal{C}$,
    there exists a substructure $G^m \subseteq G$, such that $G^m
    \equiv_m G$ and $G^m$ has\\
    $-$ atmost finitely many finite paths\\
    $-$ atmost $m$ paths which are outwardly-infinite\\
    $-$ atmost $m$ paths which are inwardly-infinite\\
    $-$ atmost $1$ path which is bidirectionally-infinite
\end{lemma}
\ti{Proof}:
    By Hanf's Theorem, there exists $t_m \in \mathbb{N}$, such that
    any two paths of length greater than $t_m$ are $m$-equivalent. For
    any graph $G \in \mathcal{C}$, define the following,

    $-$ for $i \in \mathbb{N}$, let $a^G_i$ be the number of $i$ length paths\\
    $-$ $a^G_\uparrow$ be the number of outwardly-infinite paths\\
    $-$ $a^G_\downarrow$ be the number of inwardly-infinite paths\\
    $-$ $a^G_\updownarrow$ be the number of bidirectionally-infinite paths\\
    \noindent Given $G$, consider $G^m \subseteq G$ given as,\\
    $-$ for $i \in \{0, \cdots, t_m\}$, $a^{G^m}_i = \text{min}(a_i^G, m)$\\
    $-$ $a^{G^m}_{t_m+1} = \text{min}(\sum\limits_{i=t_m+1}^\infty a_i^G, m)$\\
    $-$ for $i > (t_m+1)$, $a^{G^m}_i = 0$\\
    $-$ $a^{G^m}_\uparrow = \text{min}(a^G_\uparrow, m)$\\
    $-$ $a^{G^m}_\downarrow = \text{min}(a^G_\downarrow, m)$\\
    $-$ $a^{G^m}_\updownarrow = \text{min}(a^G_\updownarrow, 1)$

    \noindent By Hanf's theorem, it is easy to see that $G^m \subseteq
    G$ and $G^m \equiv_m G$.\epf\\

\noindent Suppose $M_1 \in \mathcal{C}'_B$ is such that $M_1 \models
\Gamma^{M_P}$. To show that $M_1 \models \phi$, we show that there
exists a substructure $M_2$ of $M_P$ such that $M_1 \equiv_m M_2$
(recall that $P$ is a model of $\phi$ and $M_P$ is the expansion of
$P$ with the elements of a chosen core $Z$ as interpretations of the
$B$ constants). Since $\phi \in PSC(B)$, $P \models \phi$, and any
substructure of $M_P$ would contain the core $Z$ of $P$, we have that
$M_2 \models \phi$. And since $M_1 \equiv_m M_2$, we would have $M_1
\models \phi$.\\ Consider the partition of $M_P$ into two parts $M_{P,
  1}$ and $M_{P, 2}$, where $M_{P, 1}$ is substructure containing all
those paths in $M_P$ which contain the interpretation of atleast one
of the constants $c_1, \cdots, c_B$ and $M_{P, 2}$ contains all the
paths in $M_P$ which are not in $M_{P, 1}$. Similarly, consider the
partition of $M_1$ into $M_{1, 1}$ and $M_{1, 2}$. There are two cases
to consider.\\

\noindent \textbf{Case 1 :} \emph{There exists a bound $n_P$ such that
  all paths in $P$ (and $M_P$) have length less than $n_P$}\\ Note
that since $M_1 \models \Gamma^{M_P}$, for every finite substructure
of $M_1$, there exists an isomorphic substructure of $M_P$. And since
all paths in $M_P$ have length less than $n_P$, we have that all paths
in $M_1$ have length less than $n_P$ as well. Consider the
substructure $M_1^S = M_{1, 1} \sqcup M_{1, 2}^m \subseteq M_1$ (where
$M_{1, 2}^m$ is as defined in Lemma \ref{lemma:m-sig}). Clearly,
$M_1^S \equiv_m M_1$. Moreover, since both $M_{1, 1}$ and $M_{1, 2}^m$
are finite, $M_1^S$ is finite, hence there exists a substructure $M_2
\subseteq M_P$, such that $M_1^S$ and $M_2$ are isomorphic. And since
$M_2 \models \phi$ (see above for the reasoning), we have $M_1^S
\models \phi$ and hence $M_1 \models \phi$ (since $M_1^S \equiv_m
M_1$).\\

\noindent \textbf{Case 2 :} \emph{There are atleast $(B + m + 2)$
  paths in $M_P$ which are infinite in both directions.}\\ Consider a
path $L$ in $M_1$ containing the interpretation $a_i$ of a constant
$c_i$. Since $M_1 \models \Gamma^{M_P}$, one can see that $L$ must be
a subpath of some path in $M_P$ - infact subpath of some path in
$M_{P, 1}$. Thus, arguing similarly for each path $L \subseteq M_{1,
  1}$, we have $M_{1, 1} \subseteq M_{P, 1}$. Also, since there are
$(B + m + 2)$ bidirectional-infinite paths in $M_P$, atleast $(m+2)$
of these would be present in $M_{P, 2}$. Now, since $M_{1, 2}^m
\subseteq M_{1, 2}$ (as defined in Lemma \ref{lemma:m-sig}) contains,
\begin{itemize}
    \item finitely many finite paths - all of these can be embedded in
      a single bidirectional infinite path
    \item atmost $m$ outwardly-infinite and atmost $m$
      inwardly-infinite paths - all of these can be embedded in $m$
      bidirectional-infinite paths
    \item atmost $1$ bidirectional-infinite path : can be embedded in
      a single bidirectional-infinite path.
\end{itemize}
it follows that $M_{1, 2}^m$ can be embedded into $M_{P, 2}$. Thus,
$M_1^S = M_{1, 1} \sqcup M_{1, 2}^m \subseteq M_{P, 1} \sqcup M_{P, 2}
= M_P$. Hence $M_1^S \models \phi$. And since $M_{1, 2}^m \equiv_m
M_{1, 2}$, we have $M_1^S \equiv_m M_1$, and hence $M_1 \models
\phi$.\\

\noindent Thus, we have shown that if $M_1 \in \mathcal{C}'_B$ and
$M_1 \models \Gamma^{M_P}$, then $M_1 \models \phi$.\epf\\

We now look at some classes of structures over which Theorem
\ref{theorem:PSC(B)=E^BA*} fails.

\section{Theorem \ref{theorem:PSC(B)=E^BA*} fails over special classes of structures}\label{section:failure-of-conjecture}

We first look at the class $\mc{F}$ of all finite structures. {\lt}
theorem fails over this class and hence so does Theorem
\ref{theorem:PSC(B)=E^BA*} (for $B = 0$).  However, we have the
following stronger result. We prove it for relational vocabularies
(constants permitted).
\begin{lemma}\label{lemma:conjecture-fails-in-the-finite}
For relational vocabularies, Theorem \ref{theorem:PSC(B)=E^BA*} fails,
over $\mc{F}$, for each $B \ge 0$.
\end{lemma}

\ti{Proof}: We refer to ~\cite{gurevich-alechina} for the
counterexample $\chi$ for {\lt} in the finite. Let $\tau$ be the
vocabulary of $\chi$ (i.e. $\{\leq, S, a, b\}$) along with a unary
predicate $U$.  Let us call an element $x$ as having colour 0 in a
structure if $U(x)$ is true in the structure and having colour 1
otherwise. Let $\varphi$ be a sentence asserting that there are
exactly $B$ elements having colour 0 and these are different from $a$
and $b$. Then consider $\phi = \neg \chi \wedge \varphi$.  Check that
since $\neg \chi$ is preserved under substructures in the finite, in
any model of $\phi$, the $B$ elements of colour 0 form a core of the
model w.r.t. $\phi$. Then $\phi \in PSC(B)$. Suppose $\phi$ is
equivalent to $\psi$ given by $\exists^B \bar{x} \forall^n \bar{y}
~\beta$ where $\beta$ is quantifier-free. Observe that in any model of
$\phi$ and $\psi$, any witness for $\psi$ must include all the $B$
elements of colour 0 (else the substructure formed by the witness
would not model $\varphi$ and hence $\phi$, though it would model
$\psi$). Consider the structure $M = (\{0, 1, \ldots, B + 2n + 3\},
\leq, S, a, b, U)$ where $\leq$ is the usual linear order on numbers,
$S$ is the (full) successor relation of $\leq$, $a = 0, b = B+2n+3$
and $U = \{1, \ldots, B\}$. Now $M \not\models \phi$ since $M
\not\models \neg \chi$.  Consider $M_1$ which is identical to $M$
except that $S(B+n+1, y)$ is false in $M_1$ for all $y$. Then $M_1
\models \phi$ so that $M_1 \models \psi$. Any witness $\bar{a}$ for
$\psi$ must include all the $B$ colour 0 elements of $M_1$. Then
choose exactly the same value, namely $\bar{a}$, from $M$ to assign to
$\bar{x}$. Choose any $\bar{b}$ as $\bar{y}$ from $M$. Check that it
is possible to choose $\bar{d}$ as $\bar{y}$ from $M_1$
s.t. $\M{M}{\bar{a}\bar{b}}$ is isomorphic to
$\M{M_1}{\bar{a}\bar{d}}$ under the isomorphism $f$ given by $f(0) =
0, f(B+2n+3) = B+2n+3, f(a_i) = a_i$ and $f(b_i) = d_i$ where $\bar{a}
= (a_1, \ldots, a_B)$, $\bar{b} = (b_1, \ldots, b_n)$ and $\bar{d} =
(d_1, \ldots, d_n)$. Then since $M_1 \models \beta(\bar{a}, \bar{d})$,
$M \models \beta(\bar{a}, \bar{b})$. Then $M$ models $\psi$, and hence
$\phi$. But that is a contradiction.\epf\\

The example expressed by $\chi$ can also be written as a sentence in a
purely relational vocabulary.  The sentence $\phi$ below is over the
vocabulary $\tau = \{ \leq, S, U\}$. We leave it to the reader to
reason out (in the same manner as in ~\cite{gurevich-alechina}) that
$\phi$ is preserved under substructures in the finite but is not
equivalent to any universal sentence.

\begin{tabular}{lll}
$~~\phi$ & = & $\chi_1 \wedge \chi_2 \wedge \chi_3$ where\\
\end{tabular}

\begin{tabular}{llll}
$\chi_1$ & = & 
$\forall x \forall y \forall z$ & $ ((x \leq x) ~\wedge 
~ ((x \leq y) \vee (y \leq x))~ \wedge$\\
& & & $~ ((x \leq y) \wedge (y \leq z)) \rightarrow (x \leq z))$
\end{tabular}

\begin{tabular}{lll}
$\chi_2$ & = & $\forall x \forall y ~S(x, y) \rightarrow \forall z (((x
\leq z) \wedge (x \neq z)) \rightarrow (y \leq z))$\\ 
$\chi_3$ & = & $\exists z \forall x_1 \forall x_2 ~(\bigwedge_{i =
  1}^{i = 2} \neg U(x_i) \wedge (x_1 \neq x_2)) \rightarrow
(\chi_4(x_1, x_2, z) \vee \chi_4(x_2, x_1, z))$\\ 
\end{tabular}

\begin{tabular}{llll}
$\chi_4(x_1, x_2, z)$ & = &
$\forall y$ & $((x_1 \leq y) \wedge (y \leq x_2) \wedge$\\
& & & $((y \neq x_1) \wedge (y \neq x_2)) \rightarrow U(y)~ \wedge$\\
& & & $((z \neq x_2) \wedge \neg S(z, y)))$ 
\end{tabular}\\

Then one can do a similar proof as above to show that for purely
relational vocabularies too, for each $B \ge 0$, Theorem
\ref{theorem:PSC(B)=E^BA*} fails over $\mc{F}$.\\

So far, in all the cases we have seen, it has always been the case
that Theorem \ref{theorem:PSC(B)=E^BA*} and {\lt} theorem either
are both true or are both false. We then finally have the following
result which is our first instance of a class of structures over which
\emph{{\lt} theorem holds but Theorem \ref{theorem:PSC(B)=E^BA*} fails}.

\begin{theorem}
Over the class $\mc{C}$ of graphs in which each graph is a finite
collection of finite \emph{undirected} paths, for each $B \ge 2$,
there is a sentence $\phi \in PSC(B)$ which is not equivalent to any
$\exists^B \forall^*$ sentence. However, {\lt} theorem holds over
$\mc{C}$.
\end{theorem}
\ti{Proof}: {\lt} theorem holds from the results of Dawar et al. over
bounded degree structures ~\cite{dawar-pres-under-ext}.  As a
counterexample to Theorem \ref{theorem:PSC(B)=E^BA*} for $B \ge 2$,
consider condition $D_1$, parametrized by $B$, which asserts that
there are atleast $B$ paths (0 length included) in the graph. We show
that this is FO definable because the following equivalent condition
$D_2$, parametrized by $B$, is FO definable: (The number of nodes of
degree 0) + $\frac{1}{2} \times$ (the number of nodes of degree 1)
$\ge B$.  We briefly explain this equivalence between $D_1$ and
$D_2$. Consider a graph satisfying $D_1$. Let $k$ be the number of
0-length paths so that there are atleast $B - k$ paths of length $\geq
1$. Each of the latter paths has exactly 2 nodes of degree 1. Then it
is easy to check that condition $D_2$ holds. Conversely, suppose a
graph satisfies $D_2$, but it has less than $B$ paths. Let $k$ be the
number of $0$-length paths so that there are atmost $B - 1 - k$ paths
of length $\geq 1$. Each of the latter paths has exactly 2 nodes of
degree 1. Then, (the number of nodes of degree 0) + $\frac{1}{2}
\times$ (the number of nodes of degree 1) $\leq (k + \frac{1}{2}
\times 2 \times (B - 1 - k)) < B$ -- contradicting $D_1$. Then $D_2$
implies $D_1$.

Then, given $B$, $D_1$ is expressible by a FO sentence $\phi$ since
$D_2$ is FO expressible (the latter is easy to see).

To see that $\phi$ is in $PSC(B)$, in any model, observe that the set
of nodes formed by picking up one end point each of $B$ distinct paths
is a core.

Now suppose that $\phi$ is equivalent over $\mc{C}$ to $\psi =
\exists^B \bar{x}~ \forall^n \bar{y} ~ \beta (\bar{x}, \bar{y})$ for
some $n \in \mathbb{N}$ where $\beta$ is quantifier-free. Consider a
graph $M$ which has exactly $\left \lceil \frac{B}{2} \right \rceil$
paths, each of length $\ge 5n$ (There is nothing sacrosanct about the
number 5 - it is just sufficiently large for our purposes). By
definition, $M \not\models \phi$ and hence $M \not\models \psi$. Label
the end points of these paths as $p_1, p_2, p_3, \ldots, p_{2\cdot k}$
where $k = \left \lceil \frac{B}{2} \right \rceil$. Now consider a
graph $N$ having exactly $B$ paths, each of length $\ge 5n$ . By
definition, $N \models \phi$ and hence $N \models \psi$. Then there
exists a witness $\bar{a} = (a_1, \ldots, a_B)$ in $N$ for
$\psi$. Observe that no two of the $a_i$s can be in the same path else
taking the substructure of $N$ formed by just the paths containing
$\bar{a}$, one would get a model of $\psi$ and hence $\phi$ - but the
number of paths in this model would be $\leq B - 1$, giving a
contradiction. We now choose points $b_1, \ldots, b_B$ in $M$ as
follows. For $i \in \{1, \ldots, B\}$, if $a_i$ is at a distance of
atmost $n$ from any end point in $N$, then choose $b_i$ to be at the
same distance from $p_i$ in $M$. Else choose $b_i$ to be at a distance
of $n$ from $p_i$ in $M$. Assigning $\bar{b} = (b_1, \ldots, b_B)$ as
$\bar{x}$, choose any $\bar{d}$ as $\bar{y}$ from $M$. Check that it
is possible to choose $\bar{e}$ as $\bar{y}$ from $N$
s.t. $\M{M}{\bar{b}\bar{d}}$ is isomorphic to $\M{N}{\bar{a}\bar{e}}$
under the isomorphism $f$ given by $f(b_i) = a_i, f(d_j) = e_j$ where
$\bar{d} = (d_1, \ldots, d_n)$ and $\bar{e} = (e_1, \ldots,
e_n)$. Since $N \models \beta(\bar{a}, \bar{e})$, $M \models
\beta(\bar{b}, \bar{d})$. Then $M$ models $\psi$ -- a
contradiction.\epf\\

\underline{\tbf{Important Note}}: For $B = 2$, the sentence $\phi$
above is equivalent to asserting that either (i) there are atleast $2$
nodes of degree exactly 0 or (ii) there are atleast $3$ nodes of
degree atmost 1. Consider the following condition for $B \ge 2$ whose
special case for $B = 2$ is the condition just mentioned: Either (i)
there are atleast $B$ nodes of degree exactly 0 or (ii) there are
atleast $B+1$ nodes of degree atmost 1. This condition, for a given
$B$, is easily seen to be expressible as a FO sentence $\xi$ (in fact,
$\xi$ is of the form $\exists^{B+1} \forall^*$). \underline{But for $B
  > 2$, $\xi \notin PSC(B)$}. To see this, consider a graph $M$
containing exactly $2$ paths $P_1$ and $P_2$ of length $\ge 1$ and
$B-3$ paths of length $0$ (the total number of paths is then $<
B$). We will show that $M$ has no core (w.r.t. $\xi$) of size atmost
$B$. Firstly, $M \models \xi$ since $M$ has $B+1$ nodes of degree
atmost 1. If $\xi \in PSC(B)$, then $M$ has a core $C$ of size atmost
$B$. There are 2 cases: (a) One of $P_1$ or $P_2$ has atmost 1 core
element. (b) Both $P_1$ and $P_2$ have atleast 2 core elements. In
case of (b), note that atleast one of the 0-length paths will not
contain any core element. Then consider the substructure $M_1$ of $M$
without this path - this contains all core elements and hence must
satisfy $\xi$. However, there are exactly $B$ elements of degree
atmost 1 in $M_1$ and hence $M_1$ violates $\xi$. In case of (a),
there are two subcases: (i) One of $P_1$ or $P_2$, say $P_1$ w.l.o.g.,
contains no core element. Then the substructure $M_1$ of $M$ which is
all of $M$, but excluding $P_1$, contains all core elements and must
hence model $\xi$. But $M_1$ contains exactly $B-1$ nodes of degree
atmost 1; so it violates $\xi$. (ii) One of $P_1$ or $P_2$, say $P_1$
w.l.o.g., contains exactly 1 core element say $a$. Let $M_1$ be the
substructure of $M$ without $P_1$. Consider the disjoint union $M_3$
of $M_1$ and the substructure $M_2$ of $M$ induced by $a$. Then $M_3
\subseteq M$ contains all core elements and must hence model
$\xi$. But $M_3$ contains exactly $B$ nodes of degree atmost 1; so it
violates $\xi$.

In all cases, we have a contradiction. Hence $M$ has no core of size
$\leq B$. Hence $\xi \notin PSC(B)$.

Interestingly however, Theorem \ref{theorem:PSC(B)=E^BA*} holds over $\mc{C}$ for $B=1$ as
we shall see in the next Lemma. We also give a simpler proof for the
case of $B = 0$ i.e. {\lt} over $\mc{C}$.

\begin{lemma}
Over $\mc{C}$, for $B \leq 1$, $\phi \in PSC(B)$ iff $\phi$ is
equivalent to $\psi$ where $\psi = \exists^B \bar{x}~ \forall^n
\bar{y} ~\phi|_{\bar{x}\bar{y}}$ for some $n \in \mathbb{N}$.
\end{lemma}
\ti{Proof}: Let the quantifier rank of $\phi$ be $m$. By Hanf's
theorem, we have the following:
\begin{enumerate}[A]
\item There exists a number $t_m \in \mathbb{N}$ such that any two
  undirected paths of length greater than $t_m$ are $m-$equivalent.
\item There exists a number $s_m \in \mathbb{N}$ such that given a
  structure $G = (P, a)$ where $P \in \mc{C}$ is (finite) path of
  length greater than $s_m$ and $a$ is a designated element of $P$,
  there is a substructure $G_1 = (P_1, a)$ of $G$ s.t. (i) $P_1$ is a
  subpath of $P$ containing the designated element $a$, (ii) $|P_1|
  \leq s_m$ and (iii) $G \equiv_m G_1$.
\item
For any graph $G \in \mc{C}$, let $a^G_i$ be the number of undirected
paths of length $i$ in $G$. Now, given graph $G \in \mc{C}$, we
consider a graph $G^m \subseteq G$ as follows (similar to the method
in the proof of Theorem
\ref{theorem:conjecture-holds-over-directed-paths}):

 $-$ for $i \in \{0, \cdots, t_m\}$, $a^{G^m}_i = \text{min}(a_i^G, m)$\\
    $-$ $a^{G^m}_{t_m+1} = \text{min}(\sum\limits_{i=t_m+1}^\infty a_i^G, m)$\\
    $-$ for $i > (t_m+1)$, $a^{G^m}_i = 0$\\

By Hanf's theorem, it is easy to verify that $G^m \equiv_m G$.
\end{enumerate}

Now consider the statement of the (current) lemma for $B = 1$.  Let $n
= s_m + \sum\limits_{i=0}^{i = t_m+1} (m \cdot (i + 1))$ and consider
$\psi$ given by $\psi = \exists x ~\forall^n \bar{y}
~\phi|_{x\bar{y}}$.  That $\phi \rightarrow \psi$ follows from Lemma
\ref{lemma:PSC(B)-sentence-implies-relativized-version}.  For the
converse, suppose $G \models \psi$. Let $a$ be a witness and let $P$
be the path in $G$ on which $a$ appears. Consider the vocabulary
$\tau_1 = \{E\} \cup \{c_1\}$ where $c_1$ is a fresh constant and
consider $\mc{G} = (G, a)$ - the $\tau_1$-structure obtained by
expanding $G$ with $a$ as the interpretation for $c_1$. Let $\mc{G} =
\mc{G}_1 \sqcup G_2$ where $\mc{G}_1 = (P, a)$ and $G_2 \in \mc{C}$ is
the collection of all paths in $G$ other than $P$. Note that we have
abused the $\sqcup$ notation slightly but the idea of separating $P$
and $a$ from the rest of $G$ is clear. Now, \\

$-$ Let $\mc{G}_1' \subseteq \mc{G}_1$ be the structure ensured by
  [B] above. Then (i) $|\mc{G}_1'| \leq s_m$ and (ii) $\mc{G}_1'
  \equiv_m \mc{G}_1$.\\
~~$-$ Let $G_2^m$ be as given by [C] above. Then (i) $G_2^m \subseteq
  G_2$, (ii) $|G_2^m| \leq \sum\limits_{i=0}^{i = t_m+1} m \cdot (i + 1) $
  and (iii) $G_2^m \equiv_m G_2$.\\

Then $\mc{G}' = (\mc{G}_1' \sqcup G_2^m) \equiv_m (\mc{G}_1 \sqcup
G_2) = \mc{G}$. Also $\mc{G}' \subseteq \mc{G}$. Note that $|\mc{G}'|
\leq s_m + \sum\limits_{i=0}^{i = t_m+1} m \cdot (i + 1) = n$. Now
since $G \models \psi$, choose $x = a$ and $\bar{y} = \bar{d}$ where
$\bar{d}$ is any tuple containing exactly the elements of $\mc{G}'$ -
this is possible since $|\mc{G}'| \leq n$ as we just saw. Then $(G, a,
\bar{d}) \models \phi|_{\bar{x}\bar{y}}$ so that $\mc{G}' \models
\phi$. Then $\mc{G} \models \phi$ and hence $G \models \phi$.

For $B = 0$, there is no $\mc{G}_1$ and hence no $\mc{G}_1'$. It is
easy to see that the same proof goes through.\epf

\section{Additional observations on {\lt} theorem over the class of all finite structures}\label{section:los-tarski-additional-observations}

We will refer to truth or failure of {\lt} over the class of all
finite structures simply as the truth or failure of {\lt} `in the
finite'.

Now as observed earlier in Sections \ref{section:success-stories} and
\ref{section:failure-of-conjecture}, while {\lt} fails in the finite,
there are special fragments of FO for which {\lt} is \emph{true} in
the finite. We present below two additional fragments of FO for which
{\lt} is true in the finite. This would follow from their
combinatorial proofs and hence we state the results below for
arbitrary structures.

\begin{lemma}\label{lemma:los-tarski-holds-for-EA}
Consider $\phi$ of the form $\exists x \forall y \psi(x, y)$ in a
purely relational vocabulary $\tau$. If $\phi \in PS$, then $\phi$ is
equivalent to $\varphi = \forall z_1 \ldots \forall z_n \phi|_{\{z_1,
  \ldots, z_n\}}$ where $n = 2^{|\tau|}$. Further, this bound is tight
i.e. there is a $\exists \forall$ sentence in $PS$ which is not
equivalent to a universal sentence with less than $n$ quantifiers.
\end{lemma}
\ti{Proof}: 

From Lemma \ref{lemma:PSC(B)-sentence-implies-relativized-version}, it
follows that if $M \models \phi$ then $M \models \varphi$. Therefore
to prove the lemma, it suffices to show that if $M \models \varphi$,
that is, every substructure of $M$ with size atmost $n$ is a model of
$\phi$, then infact $M \models \phi$. We prove it by contradiction, so
assume that $M \models \varphi \wedge \lnot \phi$. The main idea is to
use $M$ to come up with a structure which models $\phi$, but which has
a substructure which is a non-model of $\phi$. This contradicts that
$\phi \in PS$. [Note that $|M| > n$ for such an $M$, since if $|M| \le
  n$ and $M \models \varphi$ then $M \models \phi$ as well.]
  
Since every substructure of $M$ with size atmost $n$ models $\phi$,
every $1$ sized substructure of $M$ is a model of $\phi$, and hence
$\psi(x,x)$ is true for every $x \in M$ (recall that $\phi = \exists x
\forall y \psi(x, y)$).  Now note that $n = 2^{|\tau|}$ is the number
of all $1$-types possible over the vocabulary $\tau$ upto equivalence
(An \emph{$i$-type} of $\tau$ is a quantifier-free formula over $\tau$
which uses just $i$ variables. The number of $i$ types is finite upto
equivalence. See ~\cite{libkin} where our $i$-type is called
\emph{rank-0, $i$-type}). Denote the $1$-types as $\{\sigma_0, \cdots,
\sigma_{n-1}\}$, and $\sigma_i(x)$ denotes that $x$ is of $1$-type
$\sigma_i$. Suppose that there exists an element $x_0$ of $1$-type
$\sigma_i$ in $M$.  Since $M \models \forall x \exists y \lnot
\psi(x,y)$, there exists a $y_0$ such that $\psi(x_0,y_0)$ is false in
$M$.  However, since every substructure of size atmost $n$ is a model
of $\phi$, the substructure $\M{M}{\{x_0, y_0\}} \models \phi$ and
hence $\psi(y_0,x_0)$ must be true in $M$ (since either $x_0$ or $y_0$
must act as a witness for $x$ in $\phi$. But $\psi(x_0,y_0)$ is
false. Hence $x_0$ cannot be the witness). Let $y_0$ be of $1$-type
$\sigma_k$.

Suppose that it is possible to have a structure $A$ with just two
elements $\{a_0, a_1\}$ such that $\sigma_i(a_0)$, $\sigma_i(a_1)$ and
$\lnot \psi(a_0,a_1)$ hold. Then consider the structure $X$ with
universe $\{a_0, a_1, a_2, b\}$ such that (i) $\sigma_i(a_j)$ holds
for $j \in \{0, 1, 2\}$ (ii) $\sigma_k(b)$ holds

(iii) $\lnot \psi(a_j,a_{(j+1) \text{mod}~ 3})$ holds for $j \in \{0,
1, 2\}$ (iv) $\psi(b,a_j)$ holds for $j \in \{0, 1, 2\}$ and (v)
$\psi(b,b)$. Such a structure exists because all the 1-types and
2-types have been \emph{copied} from other structures, namely, (i),
(iii) are copied from A and (ii), (iv), (v) copied from $M$. Clearly,
$X \models \phi$, since $b \in X$ acts as a witness for $x$ in
$\phi$. However, the substructure of $X$ induced by $\{a_0, a_1, a_2\}
\not\models \phi$. This contradicts the given assumption of $\phi \in
PS$. Hence, it is not possible to have a structure $A$ as assumed, and
hence taking a structure $A'$ with two elements $a_0, a_1$ such that
$\sigma_i(a_0)$, $\sigma_i(a_1)$ hold, necessitates that
$\psi(a_0,a_1)$ must hold (Note that for every $1$-type $\sigma_i$ in
$M$, one can construct such an $A'_i$).

Consider $M'$ to be a substructure of $M$ which contains exactly one
element of each $1$-type present in $M$. Clearly $|M'| \le n$ and
hence $M' \models \phi$. Thus, there exists $x_1 \in M'$ such that for
every $y_1 \in M'$, $\psi(x_1, y_1)$ holds. Suppose that
$\sigma_l(x_1)$ holds. Construct an extension $\bar{M}$ of $M$ with an
additional element $z_0$ such that (i) $\sigma_l(z_0)$ holds (ii)
$\forall y \in M~ \psi(z_0,y)$ holds (iii) $\psi(z_0,z_0)$ holds. Such
a structure $\bar{M}$ exists because all the 1-types and 2-types have
been \emph{copied} from other structures, namely (i), (iii) are copied
from $M$, (ii) is copied from $M$ for $y$ satisfying $\lnot
\sigma_i(y)$, and for $y$ satisfying $\sigma_i(y)$, the 2-type is
copied from $A'_l$. Clearly, $\bar{M} \models \phi$ as $z_0 \in
\bar{M}$ acts a witness for $x$ in $\phi$. However, $M \subseteq
\bar{M}$ and $M \not\models \phi$. This again contradicts that $\phi
\in PS$. Hence, our original assumption that there exists $M$ such
that $M \models \varphi \wedge \lnot\phi$ is incorrect. Then $\varphi
\rightarrow \phi$.\\

To prove the optimality of the bound, consider the following example
over a vocabulary of $k$ unary predicates. We construct a formula
$\phi$ such that the smallest $n$ for which $\phi \leftrightarrow
\forall z_1 \cdots \forall z_n \phi|_{\{z_1,\dots,z_n\}}$ is infact $n
= 2^k$. Suppose for contradiction that $\phi \leftrightarrow \forall
z_1 \ldots \forall z_n \psi$. Then by Lemma
\ref{lemma:its-always-possible-to-replace-matrix-by-phi-itself}, $\phi
\leftrightarrow \forall z_1 \cdots \forall z_{n-1}
\phi|_{\{z_1,\dots,z_{n-1}\}}$. Let $\{\sigma_0, \cdots,
\sigma_{n-1}\}$ be the set of all $1$-types. 

Define $\phi = \exists x \forall y \ \bigwedge\limits_{i=0}^{n-1}
(\sigma_i(x) \rightarrow \lnot \sigma_{(i+1)~\text{mod}~n}(y))$. It is
easy to check that the semantic interpretation of $\phi$ implies that
$M \models \phi$ if and only if there exists atleast one $1$-type
$\sigma_j$ which is \emph{not} present in $M$. Now consider the
structure $M$ which has exactly one copy of each $1$-type
$\sigma_i$. Clearly, in every substructure of $M$ which has size less
than or equal to $n-1$, there exists atleast one $1$-type which is
missing. Hence $M \models \forall z_1 \cdots \forall z_{n-1}
\phi|_{\{z_1,\dots,z_{n-1}\}}$. However, $M \not\models \phi$ as all
$1$-types are present in $M$. This is a contradiction. Hence $\phi
\not\leftrightarrow \forall z_1 \cdots \forall z_{n-1}
\phi|_{\{z_1,\dots,z_{n-1}\}}$, and thus, the bound $n = 2^{|\tau|}$
is optimal.\epf

\begin{lemma}\label{lemma:los-tarski-holds-for-E*A-without-equality}
Let $\tau$ be a purely relational vocabulary and $\phi$ be a sentence
in $FO(\tau)$ s.t. (i) $\phi = \exists x_1 \ldots \exists x_k \forall
y \psi(x_1, \ldots, x_k, y)$ where $\psi$ is quantifier free and no
$\exists$ variable is compared with a $\forall$ variable using
equality (ii) $\phi \in PS$. Then $\phi$ is equivalent to $\varphi =
\forall z_1 \ldots \forall z_n \phi|_{\{z_1, \ldots, z_n\}}$ where $n$
is $2^{|\tau|}$.
\end{lemma}
\ti{Proof}: 

From Lemma \ref{lemma:PSC(B)-sentence-implies-relativized-version}, we
have $\phi \rightarrow \varphi$.  Therefore to prove the lemma, it
suffices to show that if $M \models \varphi$, that is, every
substructure of $M$ with size atmost $n$ is a model of $\phi$, then
infact $M \models \phi$. We prove it by contradiction, so assume that
$M \models \varphi \wedge \lnot \phi$. The main idea is to use $M$ to
come up with a structure which models $\phi$, but which has a
substructure which is a non-model of $\phi$. This contradicts that
$\phi \in PS$. [Note that $|M| > n$, since if $|M| \le n$ and $M
  \models \varphi$ then $M \models \phi$ as well.]

Consider $M'$ to be a substructure of $M$ which contains exactly one
element of each $1$-type present in $M$. Clearly $|M'| \le n$ and
hence $M' \models \phi$. Thus, there exists $a_1, \cdots, a_k \in M'$
such that for every $b \in M'$, $\psi(a_1, \cdots, a_k,b)$
holds. Construct an extension $\bar{M}$ of $M$ with $k$ additional
elements $\{z_1, \cdots, z_k\}$ such that (i) $\M{\bar{M}}{\{z_1,
  \cdots, z_k\}}$ is isomorphic to $\M{M'}{\{a_1, \cdots, a_k\}}$ via
the isomorphism $f(z_i) = a_i$ 

(ii) $\forall y \in M \ \psi(z_1, \cdots, z_k, y)$ holds. Such a
structure $\bar{M}$ exists because all $r$-types ($r \le k+1$) have
been obtained by \emph{copying} predicate values from other structures
as now explained. The types in (i) are copied from $M'$. The types in
(ii) are copied as they are from $M'$ as follows: suppose $y_0 \in M'$
has the same $1$-type as $y \in M$, then $r$-type $\{z_1, \cdots, z_k,
y\}$ in $\bar{M}$ (where $r$ is the number of distinct elements
present in $\{z_1, \cdots, z_k, y\}$) is obtained by having all
propositional statements, $\alpha(z_1, \cdots, z_k, y)$ to have the
same value in $\bar{M}$ as $\alpha(a_1, \cdots, a_k, y_0)$ in $M'$,
where there is no equality between $y$ and $z_j$ in $\alpha$. Then,
$\psi(z_1, \cdots, z_k, y_0)$ is true in $\bar{M}$, as $\psi(a_1,
\cdots, a_k, y_0)$ is true in $M'$. Also, since there are no equality
comparisons between $z_j$ and $y$ in $\psi$, $\psi(z_1, \cdots, z_k,
y)$ has the same value as $\psi(a_1, \cdots, a_k, y_0)$, even if $y_0$
was infact one of the $a_j$s itself. Thus, we have $\bar{M} \models
\phi$ as $z_1, \cdots, z_k \in \bar{M}$ act as witnesses for $x_1,
\cdots, x_k$ in $\phi$. However, $M \subseteq \bar{M}$ and $M
\not\models \phi$. This contradicts that $\phi \in PS$. Hence, our
original assumption that there exists $M$ such that $M \models \varphi
\wedge \lnot\phi$ is incorrect. Then $\varphi \rightarrow \phi$.\epf\\

We now make the following important observation given our
results. Over the class of all finite structures and for purely
relational vocabularies, the following hold:
\begin{enumerate}
\item {\lt} holds trivially for the $\Sigma^0_1$ and $\Pi^0_1$
  fragments of FO. A $\Sigma^0_1$ sentence in $PS$ is actually
  \emph{valid}. There is nothing to do in the $\Pi^0_1$ case.
\item By Lemma \ref{lemma:conjecture-for-Pi^0_2}, {\lt} holds for
  $\Pi^0_2$.
\item The counterexample to {\lt} in the finite, given as a purely
  relational sentence $\phi$ after Lemma
  \ref{lemma:conjecture-fails-in-the-finite} in Section
  \ref{section:failure-of-conjecture}, is an $\exists \forall^4$
  sentence. Then {\lt} fails in the finite for $\Sigma^0_k$ and
  $\Pi^0_k$ for all $k \ge 3$.
\item By Lemmas \ref{lemma:los-tarski-holds-for-EA} and
  \ref{lemma:los-tarski-holds-for-E*A-without-equality}, for the
  $\exists \forall$ fragment (with equality) of $\Sigma^0_2$ and the
  $\exists^*\forall$ fragment (with restricted equality) of
  $\Sigma^0_2$, {\lt} holds.
\end{enumerate}

This then leaves open only the following cases to investigate for
{\lt} in the finite for purely relational vocabularies.
\begin{enumerate}
\item Full $\exists^*\forall$ fragment (in particular, the
  `with-equality' case)
\item $\exists^*\forall^2$
\item $\exists^*\forall^3$
\item $\exists \forall^4$ without equality
\end{enumerate}

Any resolution of all these cases would give a \underline{complete
  characterization} of the dividing line in the class of prefix
fragments of FO, over purely relational vocabularies, between those
prefix fragments for which {\lt} holds in the finite and those for
which it does not!

We are currently trying to see if Lemmas
\ref{lemma:los-tarski-holds-for-EA} and
\ref{lemma:los-tarski-holds-for-E*A-without-equality} go through for
relational vocabularies too (constants permitted). If so, then
observing that the counterexample $\chi$ mentioned in the proof of
Lemma \ref{lemma:conjecture-fails-in-the-finite} is a $\exists
\forall^3$ sentence, the only cases left to investigate would be the
above cases of (1) and (2) and finally the $\exists \forall^3$
fragment without equality. With any resolution of these cases, we
would get a complete characterization of the dividing line in the
class of prefix fragments of FO, over relational vocabularies, between
those prefix fragments for which {\lt} holds in the finite and those
for which it does not.

\section{Proof of Theorem \ref{theorem:PSC(B)=E^BA*}}\label{section:proof-of-conjecture}

We first introduce some notations. Given a vocabulary $\tau$, we
denote by $\tau_k$, the vocabulary obtained by expanding $\tau$ with
$k$-fresh constants, say $c_1, \ldots, c_k$. Given a $\tau$-structure
$M$ and $k$ elements $b_1, \ldots, b_k$ from $M$, we denote by $(M,
b_1, \ldots, b_k)$, the $\tau_k$-structure whose $\tau$-reduct is $M$
and in which the constant $c_i$ is interpreted as $b_i$ for $1 \leq i
\leq k$.  Finally, for a $\tau$-structure $M$, we denote by $|M|$, the
power of $M$, i.e. the cardinality of the universe of $M$.\\

We begin with the following definition.

\begin{defn}($k$-cover)
Given a $\tau$-structure $M$, we call a set $K$ of $\tau$-structures
as a \emph{$k$-cover} of $M$ if (i) $N \subseteq M$ for each $N \in K$
(ii) the union of the universes of the elements of $K$ is the universe
of $M$ and (iii) for every atmost $k$-sized subset $S$ of the universe
of $M$, there exists an element of $K$ containing $S$. We call $M$ as
the \emph{union of $K$} and denote $M$ as $\bigcup K$.
\end{defn}

Note that given $M$, there always exists a $k$-cover of it - choose
the set $K$ above as $\{M\}$.

\begin{defn}(Preservation under $k$-covers)
A $FO(\tau)$-sentence $\phi$ is said to be preserved under $k$-covers,
if for all $\tau$-structures $M$ and all $k$-covers $K$ of $M$, if
every structure in $K$ satisfies $\phi$, then $M$ satisfies $\phi$.
\end{defn}

We will assume familiarity with the notion of saturations described in
~\cite{chang-keisler} and recall now the following theorems from
~\cite{chang-keisler} which we will use subsequently.

\begin{proposition}(A special case of Proposition 5.1.1(iii) in ~\cite{chang-keisler})\label{proposition:extending-saturations-to-expansions}
Given an infinite cardinal $\lambda$ and a $\lambda$-saturated
structure $M$, for every $k$-tuple $(a_1, \ldots, a_k)$ of elements
from $M$ where $k \in \mathbb{N}$, $(M, a_1, \ldots, a_k)$ is also
$\lambda$-saturated.
\end{proposition}

\begin{proposition}(Proposition 5.1.2(ii) in ~\cite{chang-keisler})\label{proposotion:finite-structures-are-saturated}
$M$ is finite iff $M$ is $\lambda-$saturated for all cardinals
  $\lambda$.
\end{proposition}

\begin{theorem}(A special case of Lemma 5.1.4 in ~\cite{chang-keisler})\label{theorem:existence-of-saturated-elementary-extensions}
Let $\tau$ be a finite vocabulary, $\lambda$ be an infinite cardinal
and $M$ be a $\tau$-structure such that $\omega \leq |M| \leq
2^{\lambda}$. Then there is a $\beta$-saturated elementary extension
of $M$ for $\beta \ge \lambda$.
\end{theorem}

\begin{theorem}(Lemma 5.2.1 in ~\cite{chang-keisler})\label{theorem:embedding-conditions}
Given $\tau$-structures $M$ and $N$ and a cardinal $\lambda$, suppose
that (i) $M$ is $\lambda$-saturated (ii) $\lambda \ge |N|$ and (iii)
every existential sentence true in $N$ is also true in $M$. Then $N$
is embeddable in $M$.
\end{theorem}

Putting Theorem
\ref{theorem:existence-of-saturated-elementary-extensions} and
Proposition \ref{proposotion:finite-structures-are-saturated} together
we get the following.

\begin{Corollary}\label{corollary:always-existence-of-saturated-elementary-extensions}
For every $\tau-$structure $M$, there exists a $\beta$-saturated
elementary extension of $M$ for some cardinal $\beta \ge \omega$.
\end{Corollary}

Towards our syntactic characterization, we first prove the following.

\begin{lemma}\label{lemma:characterization-over-saturated-models}
Given a finite vocabulary $\tau$, consider a $FO(\tau)$-sentence
$\phi$ which is preserved under $k$-covers and let $\Gamma$ be the set
of all $\forall^k \exists^*$ consequences of $\phi$. Then for all
infinite cardinals $\lambda$, for every $\lambda$-saturated structure
$M$, if $M \models \Gamma$, then $M \models \phi$.
\end{lemma}
\ti{Proof}:

If $\phi$ is either unsatisfiable or valid, then the result is
immediate.

Else, consider $M$ satisfying the assumptions above. To show that $M
\models \phi$, it suffices to show that for every atmost $k$-sized
subset $S$ of the universe of $M$, there is a substructure $M_s$ of
$M$ containing $S$ such that $M_s$ models $\phi$. Then the set $K =
\{M_s |S ~\text{is an atmost $k$-sized subset of the universe of M}\}$
forms a $k$-cover of $M$. Further since $\phi$ is preserved under
$k$-covers, $M \models \phi$.

Let $a_1, \ldots, a_k$ be the elements of a subset $S$ of the universe
of $M$. To show the existence of $M_s$, it suffices to show that there
exists a $\tau_k$-structure $N$ s.t. (i) $N$ is of power atmost
$\lambda$ (ii) every $\exists^*$ sentence true in $N$ is also true in
$(M, a_1, \ldots, a_k)$ (ii) $N \models \phi$. Since $M$ is
$\lambda$-saturated, by Proposition
\ref{proposition:extending-saturations-to-expansions}, $(M, a_1,
\ldots, a_k)$ is also $\lambda$-saturated. Then from Theorem
\ref{theorem:embedding-conditions}, $N$ is embeddable into $(M, a_1,
\ldots, a_k)$. Then the $\tau$-reduct of the copy of $N$ in $(M, a_1,
\ldots, a_k)$ can be taken to be $M_s$ referred to above.

We now show the existence of $N$ to complete the proof.

Let $P$ be the set of all $\forall^*$ sentences of $FO(\tau_k)$ which
are true in $(M, a_1, \ldots, a_k)$. Consider the set $T = \{\phi\}
\cup P$. Suppose $T$ is unsatisfiable. Then by Compactness theorem,
there is a finite subset of $T$ which is unsatisfiable. Since $P$ is
closed under taking finite conjunctions and since each of $P$ and
$\phi$ is satisfiable, there exists a sentence $\psi$ in $P$ s.t. $\{
\phi, \psi\}$ is unsatisfiable. Then $\phi \rightarrow \neg \psi$. Now
$\phi$ is a $FO(\tau)$ sentence while $\psi$ is a $FO(\tau_k)$
sentence. Then by $\forall$-introduction, $\phi \rightarrow \varphi$
where $\varphi = \forall x_1 \ldots \forall x_k \neg \psi[ c_1 \mapsto
  x_1; \ldots; c_k \mapsto x_k]$ where $x_1, \ldots, x_k$ are $k$
fresh variables and $c_i \mapsto x_i$ denotes replacement of $c_i$ by
$x_i$. Now note that since $\psi$ is a $\forall^*$ sentence, $\neg
\psi$ is a $\exists^*$ sentence (in $FO(\tau_k)$) and hence $\varphi$
is a $\forall^k \exists^*$ sentence (in $FO(\tau)$). Then $\varphi \in
\Gamma$ so that $M \models \varphi$. Then $(M, a_1, \ldots, a_k)
\models \neg \psi$. This contradicts the fact that $\psi \in P$.

Then $T$ is satisfiable.  By L\"owenheim-Skolem theorem, there is a
model $N$ of $T$ of power atmost $\lambda$. Since $N$ models every
$\forall^*$ sentence true in $(M, a_1, \ldots, a_k)$, it follows that
every $\exists^*$ sentence true in $N$ is true in $(M, a_1, \ldots,
a_k)$. Finally, since $N \models \phi$, $N$ is indeed as desired.
\epf

\begin{theorem}\label{theorem:characterizing-k-covers}
Given a finite vocabulary $\tau$, a $FO(\tau)$-sentence $\phi$ is
preserved under $k$-covers iff it is equivalent to a $\forall^k
\exists^*$ sentence.
\end{theorem}
\ti{Proof}:

Let $\Gamma$ be the set of all $\forall^k \exists^*$ consequences of
$\phi$. It is easy to see that $\phi \rightarrow \Gamma$.  For the
converse direction, suppose $M \models \Gamma$. By Corollary
\ref{corollary:always-existence-of-saturated-elementary-extensions},
there is a $\beta-$saturated elementary extension $M^+$ of $M$ for
some $\beta \ge \omega$. Then $M^+ \models \Gamma$. Then from Lemma
\ref{lemma:characterization-over-saturated-models}, it follows that
$M^+ \models \phi$. Since $M^+$ is elementarily equivalent to $M$, we
have that $M \models \phi$.

Then $\Gamma \rightarrow \phi$ and hence $\phi \leftrightarrow
\Gamma$. By Compactness theorem, $\phi$ is equivalent to a finite
conjunction of sentences of $\Gamma$. Since $\Gamma$ is closed under
finite conjunctions, $\phi$ is equivalent to a $\forall^k \exists^*$
sentence. \epf\\

We now prove Theorem \ref{theorem:PSC(B)=E^BA*}.\\

\tbf{Theorem \ref{theorem:PSC(B)=E^BA*}} ~Given a finite vocabulary
$\tau$, a $FO(\tau)$ sentence $\phi$ is in $PSC(B)$ iff it is
equivalent to a $\exists^B \forall^*$ sentence.\\

\ti{Proof:} We infer from Theorem
\ref{theorem:characterizing-k-covers} the following equivalences.

$\phi$ is equivalent to a $\exists^B \forall^*$ sentence iff

$\neg \phi$ is equivalent to a $\forall^B \exists^*$ sentence iff

For all $\tau$-structures $M$ and all $B$-covers $K$ of $M$, if
$\forall N \in K, ~ N \models \neg \phi$, then $M \models \neg \phi$
iff

For all $\tau$-structures $M$ and all $B$-covers $K$ of $M$, if $M
\models \phi$ then $\exists N \in K, ~ N \models \phi$\\

Assume $\phi \in PSC(B)$. Suppose $K$ is a $B$-cover of $M$ and that
$M \models \phi$.  Since $\phi \in PSC(B)$, there exists a core $C$ of
$M$ of size atmost $B$. Then by definition of $B$-cover, there exists
$N \in K$ s.t. (i) $N$ contains $C$ and (ii) $N \subseteq M$.  Then
since $C$ is a core of $M$, $N \models \phi$ by definition of
$PSC(B)$.  Then by the equivalences shown above, $\phi$ is equivalent
to a $\exists^B \forall^*$ sentence.  It is easy to see that an
$\exists^B \forall^*$ sentence is in $PSC(B)$.\epf

\section{Conclusion and Future Work}\label{section:conclusion}

For future work, we would like to investigate cases for which
combinatorial proofs of Theorem ~\ref{theorem:PSC(B)=E^BA*} can be
obtained. This would potentially improve our understanding of the
conditions under which combinatorial proofs can be obtained for the
{\lt} theorem as well.  An important direction of future work is to
investigate whether Theorem ~\ref{theorem:PSC(B)=E^BA*} holds for
important classes of finite structures for which the {\lt} theorem
holds.  Examples of such classes include those considered by Atserias
et al. in \cite{dawar-pres-under-ext}.  We have also partially
investigated how preservation theorems can be used to show FO
inexpressibility for many typical examples (see ~\cite{inexp-TR}).  We
would like to pursue this line of work as well in future.\\

{\large{\tbf{Acknowledgements}:}} We are extremely thankful to Anand
Pillay for helping us prove Theorem \ref{theorem:PSC_f=BSR} which
inspired us to go further to sharpen it and prove Theorem
\ref{theorem:PSC(B)=E^BA*} and study it in various special cases. Our
sincere thanks to Ben Rossman for giving us a patient hearing and for
sharing with us his unpublished result (Theorem
\ref{theorem:rossman-los-tarski}) which has been so relevant and
useful. Finally, many thanks to Nutan Limaye and Akshay Sundararaman
for discussions on inexpressibility proofs using preservation
theorems.

\bibliography{refs} 

\begin{thebibliography}{10}

\bibitem{gurevich-alechina}
N.~Alechina and Y.~Gurevich.
\newblock Syntax vs. semantics on finite structures.
\newblock In {\em Structures in Logic and Computer Science. A Selection of
  Essays in Honor of A. Ehrenfeucht}, pages 14--33. Springer, 1997.

\bibitem{dawar-pres-under-ext}
A.~Atserias, A.~Dawar, and M.~Grohe.
\newblock Preservation under extensions on well-behaved finite structures.
\newblock {\em SIAM J. Comput.}, 38(4):1364--1381, 2008.

\bibitem{dawar-hom}
A.~Atserias, A.~Dawar, and P.~G. Kolaitis.
\newblock On preservation under homomorphisms and unions of conjunctive
  queries.
\newblock {\em J. ACM}, 53(2):208--237, 2006.

\bibitem{buchi}
J.~R. B\"{u}chi.
\newblock Weak second-order arithmetic and finite automata.
\newblock {\em \textit{Z. Math. Logik Grundlagen Math. 6}}, pages 66--92, 1960.

\bibitem{chang-keisler}
C.~C. Chang and H.~J. Keisler.
\newblock {\em Model Theory}.
\newblock Elsevier Science Publishers, $3^{rd}$ edition, 1990.

\bibitem{dawar-model-theory-large}
A.~Dawar, M.~Grohe, S.~Kreutzer, and N.~Schweikardt.
\newblock Model theory makes formulas large.
\newblock In {\em ICALP}, pages 913--924, 2007.

\bibitem{gurevich-shelah}
Y.~Gurevich.
\newblock Toward logic tailored for computational complexity.
\newblock In {\em COMPUTATION AND PROOF THEORY}, pages 175--216. Springer,
  1984.

\bibitem{libkin}
L.~Libkin.
\newblock {\em Elements of Finite Model Theory}.
\newblock Springer, 2004.

\bibitem{rosen}
E.~Rosen.
\newblock Some aspects of model theory and finite structures.
\newblock {\em Bulletin of Symbolic Logic}, 8(3):380--403, 2002.

\bibitem{rossman-hom}
B.~Rossman.
\newblock Homomorphism preservation theorems.
\newblock {\em J. ACM}, 55(3), 2008.

\bibitem{rossman-los-tarski}
B.~Rossman.
\newblock \textit{Personal Communication}.
\newblock 2012.

\bibitem{inexp-TR}
A.~Sankaran, N.~Limaye, A.~Sundararaman, and S.~Chakraborty.
\newblock \textit{Using Preservation Theorems for Inexpressibility Results in
  First Order Logic}.
\newblock Technical report, 2012.
\newblock URL : http://www.cfdvs.iitb.ac.in/reports/index.php.

\end{thebibliography}
\bibliographystyle{plain}

\end{document}